\documentclass[final,3p,times,twocolumn,authoryear]{elsarticle}

\usepackage{amssymb}
\usepackage{amsmath}
\usepackage{mathptmx}
\usepackage{overpic}
\usepackage{subfigure}
\usepackage{color}
\usepackage{mathtools}   
\usepackage[subnum]{cases}
\usepackage{xfrac}

\journal{Journal of Theoretical Biology}

\begin{document}

\begin{frontmatter}



\title{A biophysical model of cell evolution after cytotoxic treatments: damage, repair and cell response}


\author[IEMN,IBL]{M. Tomezak}

\author[IBL,USTL]{C. Abbadie}


\author[COL]{E. Lartigau}

\author[IEMN,USTL]{F. Cleri\fnref{fn1}}
\ead{fabrizio.cleri@univ-lille1.fr}

\address[IEMN]{Institut d'Electronique Microelectronique et Nanotechnologie (IEMN), UMR Cnrs 8520, 59652 Villeneuve d'Ascq, France}

\address[IBL]{Institut de Biologie de Lille (IBL), UMR Cnrs 8161, 59019 Lille, France}

\address[COL]{Centre de Lutte contre le Cancer "Oscar Lambret", 59000 Lille, France}

\address[USTL]{Universit\'e de Lille I, Sciences et Technologies, 59650 Villeneuve d'Ascq, France}

\fntext[fn1]{Corresponding author.}

\begin{abstract}
We present a theoretical agent-based model of cell evolution under the action of cytotoxic treatments, such as radioteraphy or chemoteraphy. The major features of cell cycle and proliferation, cell damage and repair, and chemical diffusion are included. Cell evolution is based on a discrete Markov chain, with cells stepping along a sequence of discrete internal states from 'normal' to 'inactive'. Probabilistic laws are introduced for each type of event a cell can undergo during its life cycle: duplication, arrest, apoptosis, senescence, damage, healing. We adjust the model parameters on a series of cell irradiation experiments, carried out in a clinical LINAC at 20 MV, in which the damage and repair kinetics of single- and double-strand breaks are followed. Two showcase applications of the model are then presented. In the first one, we reconstruct the cell survival curves from a number of published low- and high-dose irradiation experiments. We reobtain a very good description of the data without assuming the well-known linear-quadratic model, but instead including a variable DSB repair probability, which is found to spontaneously saturate with an exponential decay at increasingly high doses. 
As a second test, we attempt to simulate the two extreme possibilities of the so-called 'bystander' effect in radiotherapy: the 'local' effect versus a 'global' effect, respectively activated by the short-range or long-range diffusion of some factor, presumably secreted by the irradiated cells. Even with an oversimplified simulation, we could demonstrate a sizeable difference in the proliferation rate of non-irradiated cells, the proliferation acceleration being much larger for the global than the local effect, for relatively small fractions of irradiated cells in the colony.
\end{abstract}


\end{frontmatter}

\parskip 2pt
\section{Introduction}
\label{sect1}

The development of cancer in a living organism follows complex paths, not without seemingly contradictory features. From the broad point of view of systems theory, the emergence of cancer cells appears as a stochastic process, in which one single cell suddenly changes its nature without a traceable cause-effect mechanism, and starts proliferating abnormally. On the other hand, the rapid and uncontrolled development of tumoral tissues is unlikely to be described as a purely stochastic phenomenon, and metastatic propagation is drastically far from a simple diffusional flow: these features rather have the character of self-organising, non-equilibrium dynamical systems, susceptible of taking on a chaotic or avalanche pattern under the effect of a small perturbation. 

Introducing the language of theoretical physics in the domain of cancer may seem unusual. However, physical-mathematical models have already accumulated a considerable tradition in cancer studies. Early analytical models based on coupled partial differential equations 
(see, e.g., \citet{brunton,sachs}), have been accompanied in recent years, and often superseded by complex numerical simulations models \citep{edelman,deisboeck,lowengrub,tracqui}, which attempt at following the space- and time-dependent dynamics of cancer growth, by adding an increasing wealth of details and phenomenological correlations coming from biochemical and clinical studies.

\color{black} Despite the considerable efforts in modelling, cancer treatments are still relying on a substantially empirical knowledge. \color{black} Radiotherapy employs ionising radiation to eradicate cancer cells, mainly through the generation of DNA double-strand breaks (DSB), although the detailed mechanisms by which DSB and other sub-cellular lesions are generated are still quite far from clear. Empirical descriptions, such as the so-called linear-quadratic model \citep{dale} are still extensively used in radiotherapy, to describe cell damage upon the delivery of ionising radiation, together with extensions, such as the 'Tumor Control Probability' model \citep{kutcher}, aimed at predicting the clinical efficacy of radiotherapeutic protocols. However, a detailed correlation between the radiation dose and its microscopic outcomes, both at the cell and tissue level, is still missing. 

In this work we develop, implement, calibrate, and apply a discrete-cell model with internal degrees of freedom, describing both normal and abnormal cell evolution, and accounting for localized damage and repair mechanisms, with the aim of studying the long-term evolution of a cell population subject to cytotoxic therapeutic treatments. Our main interest and application concerns the immediate and delayed action of radiotherapy treatments. However, the formalism developed here is enough general to be easily applicable to other cytotoxic agents, such as chemotherapy, oxidative poisoning, environmental (UV) radiation damage, and so on. 

The virtual cell population is represented by a large assembly (up to several millions) of individual stochastic agents (see e.g. \citet{byrne,wang,cilfone}), endowed with a number of probabilistic properties (phenotypes), which allow to follow the evolution of individual cells through their daily cycle over very long time scales (days, months, up to years). Each cell has a local clock which goes through the G1, S, G2 and M phases, typically (but not necessarily) following a 24h cycle. Cell duplication with inheritance is allowed, with individual probability laws depending on the cell state at any given time. Normal, stem, or tumor cells of various kinds can be included. In the simplest implementation, adapted to mimicking \emph{in vitro} experiments on cell colonies, simulated cells live on a two-dimensional fixed square grid and can migrate by vicinal displacements. Diffusion of, e.g., oxygen, nutrients, or other chemical species is allowed on the same grid.

These model cells can absorb a number of different damaging events, described by a Markov chain which changes the state of each cell from healthy, to progressively damaged, to inactive. In this first paper we model only radiation-induced DNA damage in the form of  single-strand and double-strand breaks. However, other DNA lesions, such as base excision, polymerisation, clustered defects, could be included by extending the model, as well as damage to other vital cell components, such as mitochondria. Repair mechanisms are included aside of the damage, by assigning to each cell a set of probabilities to move its individual state upwards in the Markov chain state. Cells can also exit to a quiescent, or a senescent state, which then follow special paths.


Such a model will have predictive capability, after being carefully calibrated on known evolution patterns of real biological cell lines. It should be able to predict the long term evolution (ranging from week-months, up to 5-10 years time, \color{black} well-beyond the time scales accessible to direct biological experimentation) \color{black} of a population of cells with rather arbitrary characteristics, 
eventually subject to a series of time-defined treatments, for example by radiological or chemical agents, which may alter the cell vital cycle, notably by modifying its health condition and repair capabilities. In the present work, in order to calibrate the probability of inducing a SSB or a DSB, we performed photon-beam irradiation experiments in a clinical LINAC, on cultures of normal human dermal fibroblasts. The DNA damage was quantitatively analyzed by searching foci of XRCC1 and 53BP1, two proteins involved in the repair of SSBs and DSBs respectively, by means of immunofluorescence.

While biophysical models covering some of the above characteristics have been already introduced in the literature, based on various mathematical approaches (see e.g. \citet{sanchez,stewart,torquato,deisboeck,wang}, and references therein), our model aims at assembling the most relevant features, in the attempt to develop a realistic platform for the virtual modelling of the long-term evolution of cell proliferation and damage, following various types of therapeutic treatments. \color{black} It is worth noting that the coupling of agent-based models with radio- or chemio-therapy, to simulate the action of external agents on cancer growth and/or arrest, is not yet fully developed. Even the most recent attempts in this direction (see e.g. \citet{kempf,powa}) did not include an explicit simulation of the radiation damage, but rather assumed a pre-existing damage model (such as the linear-quadratic, etc.). An original contribution of the present simulation model is the introduction of explicit damage accumulation and repair, at the single-cell level. \color{black}

After an ample Section 2 describing the key details of the model, in Section 3 we include two benchmark applications, aimed at demonstrating some key features of the computer model, namely: the reproduction of cell survival curves from irradiation experiments, and a simulation of the so-called 'bystander' effect. \color{black} Note that, for the purpose of this first work, the two applications must be intended only as test cases, with no presumption of going in depth into the complex biophysical foundations, nor the medical implications of the corresponding phenomena. \color{black} 


\section{Model}
\label{sect2}

We use an agent-based model to describe the evolution of a cell population, under normal conditions, or in response to a physical or chemical perturbation. Each cell is represented as an 'agent', endowed with probabilistic rules to follow in the course of the simulation, both for their behaviour as independent units, and in interaction with each other. In our model, agents live on a fixed lattice (in the present study simply two-dimensional (2D) with fourfold symmetry), and can move on the lattice sites, carrying all the information about their state. A cell $u$ at a site $i$ will be indicated as $u(i)$. The time-dependent behaviour of each cell is characterised by a number of descriptors, collected in a state vector $\textbf{n}(t)$ (see below). Such an implementation is different from other lattice-based models, e.g. of Potts or lattice-gas type (for a review see \citet{book1}), in that the properties are dynamically carried by the agents, and not statically attributed to the lattice sites. We believe such a setting to be more flexible in the numerical implementation, notably in view of the future developments of the model.

\begin{figure*}[t]
\begin{center}
\includegraphics[scale=0.6]{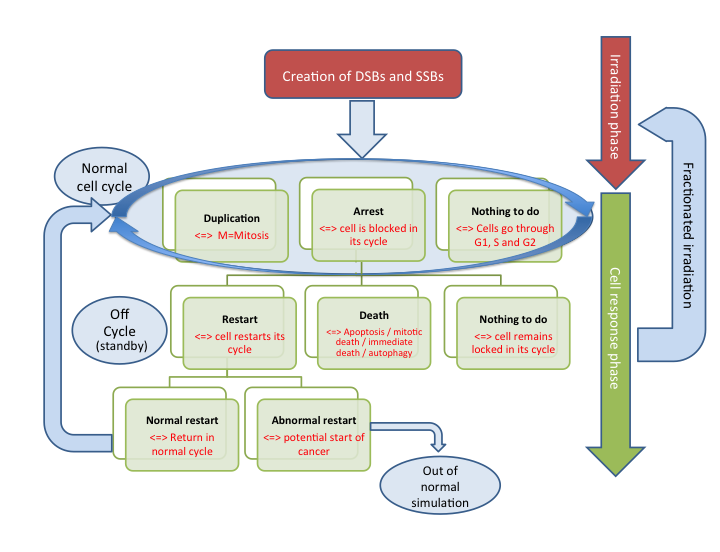}
\caption{Flowchart of the cell decision algorithm and the Monte Carlo simulation. Black lettering indicates the actions of the computer code; red lettering indicates the corresponding biological/phenotype outcomes; in white lettering, the explanatory notes. The irradiation phase produces cell damage. In the subsequent evolution stage, while cells loop in the normal 24h cycle, they can repair damage, while following their doubling pattern. In this stage cell arrest can occur because of excess damage or senescence, followed by cell death; under given conditions restart can occur that, in turn, can yield back a normal cell, or evolve into a neoplastic cell. The coupled irradiation-evolution stages can be repeated cyclically, to simulate fractionated irradiation treatment.}
\label{fig:flux}
\end{center}
\end{figure*}

Cells-agents may receive signals and input both from the environment and their neighbouring agents, as well as transmit signals to the environment and their neighbours. Agents make decisions based on both their inputs and internal state. In future developments, cells will also include layers of subcellular decision-making rules; however, here we will only consider a constant set of rules, valid for the ensemble of cells. An agent may change state, proliferate, or undergo apoptosis or necrosis in response to surrounding conditions. Cellular division requires space and a sufficient level of nutrients and oxygen. A cell may enter into a quiescent state, if the space is restricted (confluence), or if the nutrient supply is scarce. After cytotoxic treatments, a cells can arrest its life cycle. Quiescent, hypoxic or arrested cells can undergo apoptosis with a finite probability, after a defined length of time. \color{black} Figure \ref{fig:flux} illustrates a flowchart of the cell phenotype decision process. \color{black} 

\subsection{Normal cell cycle}

Aside of the global simulation time, hereafter indicated by $t$, each cell has a local clock $t_d=t-t_0$, with $t_0$ being the time of its last duplication, running over a cycle of 24h. The local time $t_d$ spans the four typical phases of the cell cycle, namely: G1 from $t_d$=0 to 750 min; S from $t_d$=751 to 1250; G2 from $t_d$=1251 to 1370; M from $t_d$=1371 to 1440 min. (The timings are attributed conventionally.) Cells can be synchronised if needed, but the typical starting configuration is obtained with all cells randomly distributed among the four phases. Cell duplication should occur in the phase M, however a degree of randomness is allowed, by introducing a duplication probability:
\begin{equation}
P_{dupl} = \frac{\phi}{4} \left( \frac{1}{ 1 + e^{-(t_d-1370)/\tau} } \right)
\label{equation:pdupl}
\end{equation}

\noindent with the duplication time constant $\tau \simeq $30 min, and $\phi=1,2,3,4$ for G1, S, G2, M, respectively.  \color{black}  Fluctuations in the duplication time of typical fibroblast cells, as well as for several other cell types including bacteria, are often described by a "shifted-gamma" probability distribution (see e.g. \citet{kutalik,rubinow,stukalin}), whose corresponding integrated cumulative distribution function is very close to a Fermi function. We use the latter in Eq.(\ref{equation:pdupl}) only because it is mathematically easier, however the numerical difference with the shifted-gamma is practically irrelevant. The dependence on the parameter $\phi$ is also rather irrelevant numerically, since in all cases the probability function (\ref{equation:pdupl}) is practically equal to zero, until $t_d$ is very close to 1370 min. (begin of M phase), around which time $P_{dupl}$ rapidly goes to 1. \color{black}

The new (daughter) cell is spatially situated next to the old one, either in an empty lattice site, or by shifting nearby cells to empty sites in order to make room for the new one. When a cell duplicates, both its local clock and that of the daughter cell are reset to $t_0$=$t$. It is possible for a cell to arrest duplication by entering in the G0 quiescent phase, for which $\phi$=0. 


Normal cells may also enter a 'senescent' state (distinct from G0) some time after their development. Among other phenotype changes, senescence modifies or suppresses the cell duplication capability, while conserving most of its metabolic activities \citep{hayflick,cristof}, \color{red} This can occur independently on the state of damage, in fact also normal, non-irradiated cells can naturally undergo senescence, under specific conditions [Refs Corinne] \color{black}. This feature is taken into account by counting the number of duplications for each cell, $D$, and introducing the maximum duplication number $D_s$ after which a cell enters into senescence, and the number $D_0$ at which the population starts having senescent cells. Then, the probability $P_{dupl}$ is multiplied by a factor $S$=1 for $D \leq D_0$, $S$=0 for $D>D_s$, and:
\begin{equation}
S = 1 - \frac{D-D_0}{D_s - D_0}
\label{equation:sene}
\end{equation}

\noindent for $D_0 < D \leq D_s$. The values of the constants $D_0$ and $D_s$ are to be adjusted so as to reproduce typical senescence rates. \color{red}Without loss of generality, in the following we will adopt $D_0$=30 and $D_s$=60, a value assessed for the case of human skin fibroblasts [Refs Corinne]. \color{black} The form of Eq.(\ref{equation:sene}) suggests that any cells that proliferated the maximum number of times $D_s$ are arrested, therefore the 'senescent' state may seem redundant with the G0. However, the G0 can be entered at any time (for example because cells attained confluence), and moreover the senescent state, notably in the case of keratinocytes, can be transient. Therefore, keeping this state allows cells to restart into a novel state even at (much) later times. \color{black}

\subsection{Diffusion}

Chemical species $\mu$ with varying concentrations $c^{\mu}$ are allowed to diffuse on the lattice sites, and are accumulated at each time step $\Delta t$ in each cell $u(i)$, instantaneously located on the lattice site $i$, according to deterministic gradient flow (Fick's law):
\begin{equation}
c^{\mu}_{u(i)} = - \sum_{j'} (c^{\mu}_{u(j')} - c^{\mu}_{u(i)} ) \frac{\Delta t}{\theta^{\mu}} + s^{\mu}_{u(j)}
\label{equation:fick}
\end{equation}

\noindent where the sum runs only on the sites $j'$ nearest neighbours of site $i$, according to the chosen lattice topology. Since the neighbours realize in this first implementation a 2D diamond topology, the right-hand side of the previous equation is just a discretized representation of the Laplacian, multiplied by the time step, with the diffusion time-scale $\theta^{\mu}$ playing the role (with appropriate dimensions) of a nominal diffusion coefficient.

A source $s^{\mu}$ can be located at one specific lattice site, or an ensemble of sites, and diffuse through the empty lattice until reaching the cells. Or, it can be contained in a cell (for example, a secreted factor), diffuse on the lattice, and move along with the displacement of the cell. 

The cell membrane represents a semi-permeable barrier to almost all molecules and ions, with permeability coefficients much smaller than the diffusion in the surrounding fluid phase. Except special cases, therefore, diffusion on empty sites is  considered instantaneous, while the membrane crossing of species $\mu$ is characterised by a diffusion time $\theta^{\mu}$ depending on the cell state and local density. 

It is worth noting that the local concentration of a species is accumulated in the cell $u(i)$ occupying the site $i$, and not on the site itself. Concentrations can build up from lattice-diffuse sources as well as from cells. For example, oxygen concentration at cell $u(i)$ is the result of the concentration field existing at site $i$, plus the eventual gradient coming from the neighboring cells $u(j')$. In the special case of oxygen diffusion, the concentration at cells deep inside the colony can be considerably smaller than that of cells at the periphery, exposed to a much higher oxygen flux. Similar examples could be the highly variable metabolic rate of cancer cells as a function of their distance from blood vases, or the complex chemistry of radiation-induced "bystander" effect; this latter will make for an application example in the following Sect. 3.

\subsection{Radiation damage}

Ionising radiation tracks deposit their energy in the cell in a time of $\sim$10$^{-13}$ s. This energy density, in the form of secondary radiation, produces a variable ionisation density on its way, over a time of $\sim$10$^{-9}$ s, and over a length which, depending on the primary radiation nature and energy, can range from a few $\mu$m to several cm. The correlation between radiation energy delivery and ionisation density is the linear energy transfer (LET). High-LET radiation, such as protons or alpha particles, have a high ionisation density along their path, however running along very straight tracks. Conversely, low-LET radiation, such as photons or low energy electrons, produce a much lower density of ionisation events per unit length, however their path is very diffuse and random, therefore their energy deposition is more homogeneous in the cell volume. Anyway, only a relatively small fraction of the actual damage is produced by \emph{direct} ionisation events, while the largest part is due to the secondary chemical species produced by the radiolysis of the molecules in the cytoplasm, i.e. mostly water, which thereby liberates free radicals OH$^\bullet$ and H$^\bullet$, as well as free electrons (which go quickly into a solvated state). Such highly reactive species diffuse and attack the DNA (\emph{indirect} damage), producing breaks in one (SSB) or both (DSB) the phosphate backbones, over a time scale of $\sim$10$^{-6}$--10$^{-3}$ s. 

The ensemble of such events in our model is considered instantaneous, and gives rise to stochastic damage events in the population of cells, according to rules which will be detailed in the following. The minimum time scale we consider is of seconds, for the irradiation time, and minutes for the follow up. Irradiation can take place according to different protocols, however it is usually delivered in batches of several tens of seconds at most. It is known that cell repair enzymes start working at considerably longer times \color{black} (see e.g. \citet{plosdsb,calini}, \color{black} therefore we will consider that no cell repair activity takes place during irradiation.

Radiation events are thought to follow a Poisson process \citep{keller}, therefore it may be justified to describe the induced damage (both direct and indirect) as a Markov chain \citep{albri,hanf}. We assume that cells can be in any state $n \in [0,m]$, with $n=0$ corresponding to a healthy cell with zero accumulated damage, and $n=m$ to a cell with a maximum of accumulated damages. The $n$-th state of the cell is described by a set of indicators, or state vector, $\textbf{n}=\{i, t_d, \phi, \lambda, z_{\nu},p_{\nu},r_{\nu},c^{\mu},\theta^{\mu}\}$, for the $\nu=1,...,k$ different types of lesions (in the present case, it is only $\nu$=1 for DSBs and $\nu$=2 for SSBs); $i$ is the lattice site occupied at time $t$; $\phi$ is the cell phase; $\lambda$ is the cell state, an index 1,2,3,... for 'normal', 'arrested', 'dead', 'neoplastic' (plus additional labels for stem, cancer cells, and so on); $z_{\nu}$ is the number of accumulated damages of type $\nu$; $p_{\nu}$ is the corresponding damage probability; $r_{\nu}$ the repair probability; $c^{\mu}$ and $\theta^{\mu}$ the concentration and diffusion time of species $\mu$.

Let us define the probability $P_n(t)$ that a cell is found in the state $n$ at time $t$, with $P_n \geq 0$ and $\sum_n P_n =1$. The equation for a continuous-time Markov chain is:
\begin{equation}
\frac{dP_n(t)}{dt} = \sum_l W_{nl} P_l(t) + S_n
\label{equ:kolm}
\end{equation}

\noindent expressing the fact that the probability of observing the state $n$ of the cell at time $t$ is given by the sum over all possible probabilities of coming from any state $l \ne n$, multiplied by the transition matrix $W_{nl}$. $S_n$ is a "source" term, representing the probability that a cell is arriving in state $n$ from a cell division at time $t$. The transition matrix sums up all the processes leading to cell state evolution:
\begin{equation}
W_{nl} = \sum_{\nu} \left( \dot{\Delta} L^{\nu}_{nl}+R^{\nu}_{nl} \right)
\end{equation}

\noindent where $L^{\nu}_{nl}$ represents the probability of making $l$--$n$=$d$ lesions of type $\nu$, induced by a dose/unit time $\Delta$ (the dot indicating time derivative), and $R^{\nu}_{nl}$ the probability of repairing $d$ lesions of type $\nu$.

When considering that the number of radiation events per unit time in each cell nucleus is relatively small, the matrix elements of $D$ can be assigned the form of a Poisson's distribution. By assuming an energy spectrum of the radiation $f(\epsilon)$, and by following the reasoning of \citet{sachs}, for
each type of damage $\nu$ the probability of going from a state with $n$ lesions to a state with $n$+$d$ lesions is written in matrix form as:
\begin{equation}
L^{\nu}_{n,n+d} = \left( \begin{array}{ccccc}  \mu_0 & 0 & 0 & ... & 0 \\
\mu_1 & \mu_0 & 0 & ... & 0 \\
\mu_2 & \mu_1 & \mu_0 & ... & 0 \\
... & ... & ... & ... & 0 \\
\mu_{d} & \mu_{d-1} & \mu_{d-2} & ... & \mu_0 \\
\end{array}  \right)
\end{equation}

\begin{equation}
\mu_k = \int_0^{\infty} \frac{e^{-\epsilon} \epsilon^k}{k!} f(\epsilon) d\epsilon
\end{equation}

\begin{figure}[t]
\begin{center}
\includegraphics[scale=0.38]{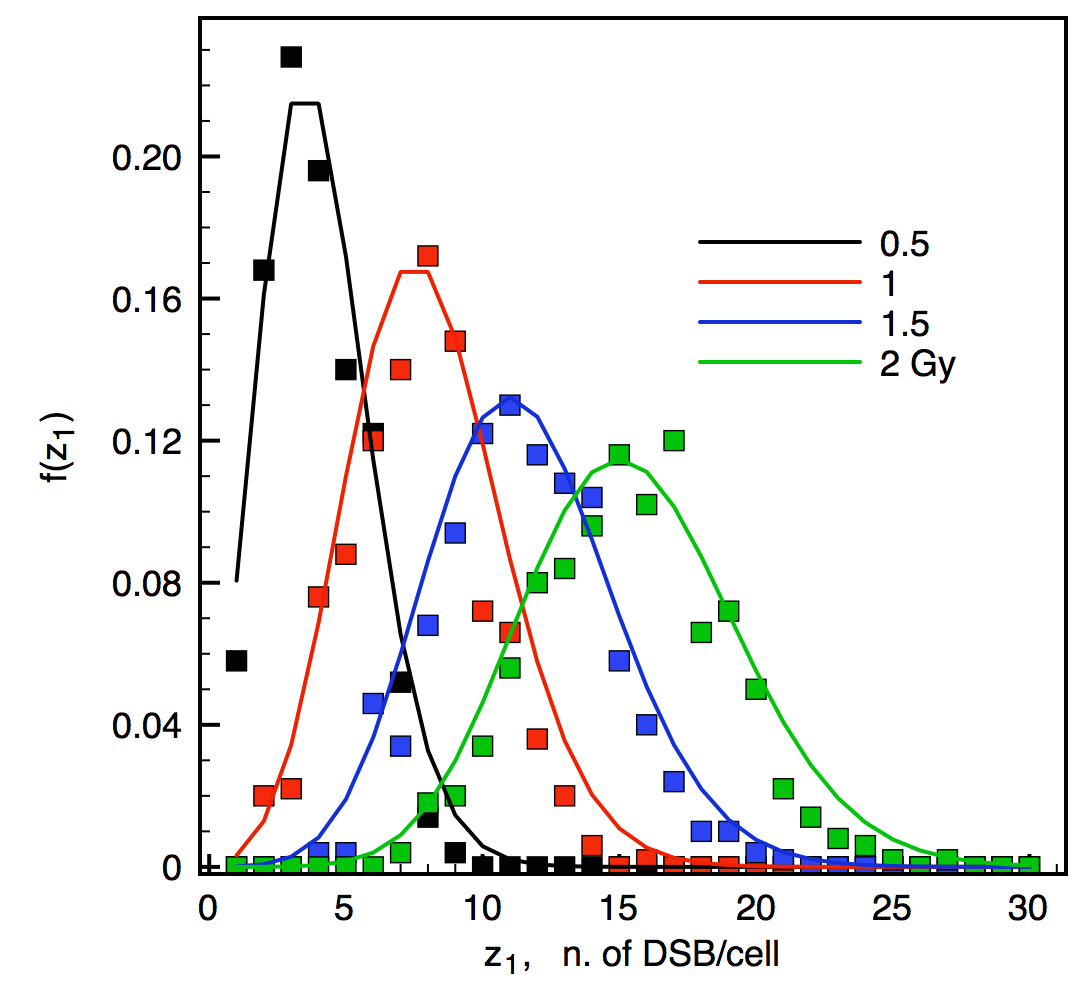}
\caption{Plot of the fraction of accumulated number of DSB lesions/cell, $z_1$, at different average dose levels from 0.5 to 2 Gy. Total number of cells, $N_c$=500. Symbols represent the raw results of the Monte Carlo simulation; continuous curves represent the fit with a Poisson probability law.}
\label{fig:prob2}
\end{center}
\end{figure}

For numerical efficiency, however, the energy spectrum is discretized, and for each energy interval the Poisson distribution is replaced by a limiting binomial \citep{keinj}:
\begin{equation}
L^{\nu}_{n,n+d} = \lim_{m_0 \to \infty} \binom{m_0-n}{d} p_{\nu}^d (1-p_{\nu})^{m_0-n-d}
\end{equation}
with $p_{\nu}=p_{\nu}(\epsilon)$ a piecewise, energy-dependent damage probability. By performing test simulations of irradiation at 2 Gy, we observe that already for $m_0 \approx 5 m$ the binomial is practically superposed to a Poisson curve for all values of dose (see Figure \ref{fig:prob2}, showing the distributions of average DSB/cell as a function of the dose).

The choice of the repair probability matrix $R$ depends on the details of the enzymatic processes leading to strand rejoining after SSB, DSB, etc. In this first paper, we take a simplistic approach by choosing a simple linear model, in which the repair fraction is dependent on the number of lesions still present, with a generic repair probability $r$, \color{black} whose numerical value is different for each type of lesion: \color{black}
\begin{equation}
R^{\nu}_{n,n-d} = -r_{\nu}d
\label{equation:exprob}
\end{equation}

\noindent which will result in an exponential repair probability distribution. Of course, there would be little difficulty in including more sophisticated models of repair kinetics (see, e.g., \citet{cucinotta}).  
 
 
 At this stage, our description of the relative biological efficiency (RBE) of the radiation is very simplified. To name a few important ingredients, we ignore complex damage types such as clustered defects, interactions between defects, evolution of one defect type into another. However, such features can be added in further developments of the numerical scheme; we are omitting them at present only to make for a more clear, albeit simplified, analysis of the results.
 
The probabilities $p_{\nu}$ and $r_{\nu}$ can be made dependent on the state of the cell. The damage probability can change according to the different radiosensitivity of a cell, e.g. increase because of an increased oxygen concentration. \color{black} The current version of the model includes the possibility of decreasing the damage probabilities $p_{\nu}$ by a simple sub-linear dependence on the total accumulated dose. However, the radiation resistance is affected by many different competing factors, most notably a low oxygen level in the tumour mass \citep{graeber}. Therefore, a more complete description of such metabolic reactions should be accounted for by means of appropriate parameters, controlled by a set of differential equations (see e.g. \citet{wang}). The repair probability may be made variable as well, for example the recruitment of repair proteins could be slowed down in a heavily damaged cell nucleus. Note that, for the sake of simplicity, in the examples of Sect. 3 we will only use constant damage and repair probabilities. \color{black}

In most experiments (see also Sect. 2.6 below) it is observed that similar cells, even belonging to the same line and culture, display variable radiosensitivity and repair capability. To mimic this effect, 
a specific dependence of the repair capability on the cell phase is included (see e.g. \citet{diko}), by setting:
\begin{equation}
r_1=
\begin{cases} \phi r & \text{if } \phi \leq 2 \\ \\
                                  \tfrac{2}{3} r    & \text{if } \phi > 2  
\end{cases} 
\label{equation:rpha}
\end{equation}

\noindent for DSB repair kinetics, \color{black} based on the fact that once the amount of DNA starts to increase in the cell nucleus before mitosis, the cell repair machinery becomes quantitatively less effective (for simplicity we take $r_2$=$const$ for SSBs, since they will not be relevant in the following simulation examples). \color{black} It is worth noting that for cancer cells of various types, the situation is more complex than for normal cells, in that a prevalence of radiosensitivity on the initial number of DSBs is more often observed, next to a reduced or even null dependence on the repair capability \citep{awadi}.

\subsection{Monte Carlo simulation}

The simulation space is defined by a 2D square lattice of $N\times N$ sites, uniquely labelled by a positive integer $i \leq N^2$. In the present work we adopt a minimal (also called 'Von Neumann') neighbourhood relationship, namely each site $i$ interacts with the four neighbours (empty or occupied) located immediately above, below, left and right, in the 2D square topology. 

An initial number of cells $N_c$, typically quite smaller than the ensemble of lattice sites, is dispersed on the lattice, either at random, or in one or more compact colonies. Cells can move on the lattice by discrete jumps to neighbouring sites. It is already known from previous studies that the underlying lattice topology may induce unrealistic features in the tissue morphology. However, in the present work cell displacements are either not considered at all, or allowed only in order to attain confluence; therefore this will not be a crucial limitation. On the other hand, if one is interested in simulating tissue and tumour morphogenesis, a denser lattice topology or a Voronoi tessellation may be adopted \citep{moreira}, and eventually a 3D extension of the model becomes necessary.

\color{black}In the present version of the computer code implementing the agent-based model, \color{black}  different types of cells can be defined according to their phenotypes. Compared to \emph{normal} cells, one could define \emph{stem} cells, which follow a peculiar duplication pattern in that a stem cell divides into a normal and another stem cell; and \emph{tumour} cells, which are defined according to a special set of the state vector \textbf{n}, e.g. an accelerated duplication probability. \color{black}In the examples to follow in Section 3, we will only use either normal cells receiving a homogeneous dose of radiation, or a mixture of irradiated vs. non-irradiated cells; two types of defects (DSB and SSB) will be tracked, although the cell death probability will be depending only on the number of accumulated DSBs. Such properties of the cell population can be easily modified, according to the subject of study.\color{black}



Solving analytically the Kolmogorov, forward-type equation (\ref{equ:kolm}) for a fully coupled set of space- and time-dependent probabilities is practically impossible. We therefore make recourse to a Monte Carlo stochastic method of solution (see Figure \ref{fig:flux}). At each discrete update by $\Delta t$ of the global time $t$, all the cells are scanned, and their state vector \textbf{n} is updated. Probabilities for the various events a cell may undergo are sampled according to a rejection technique, i.e., a random number $\xi \in (0,1)$ is drawn from a flat probability distribution, and compared to the event probability $P$. The event is accepted if $\xi < P$, otherwise it is rejected. \color{black} For numerical efficiency,  all such events are described in the following by continuous probability functions, however it may be noted that the functional shape practically corresponds to binary, "yes/no" options, at each Monte Carlo sampling step. \color{black}

For example, let us focus on a particular cell $i \in N_c$: as long as the time $t$ increases, its local clock $t_d$ advances, and its duplication probability $P_{dupl}$ increases from nearly 0 in the G1 phase, to nearly 1 in the M phase. Correspondingly, at each time step $t$, the new 0$<\xi<$1 will stochastically sample the increasing probability $P_{dupl}$, by producing a duplication event at a random time $t_0$ distributed according to $P_{dupl}$. 
The same random sampling happens for the probability of going from $z_{\nu}$ to $z_{\nu}+d$ lesions (with $\nu$=1 for DSBs and $\nu$=2 for SSBs, and damage probability $p_{\nu}$ if $d>0$, or repair probability $r_{\nu}$ if $d<0$); for the probability of going into, or escaping from, the quiescent state G0; and so on.

In a typical simulation of an irradiation cycle, cells are irradiated for some time of the order of 10s to 100s of seconds, with a time-step $\Delta t$=1 s. Then, the evolution of the partly damaged population is followed for a time of the order of several hours, with a typical $\Delta t$=60 s or larger. Eventually, the irradiation-evolution cycle may be repeated, to simulate fractionated radiotherapy. For the sake of simplicity, we assume that no cell evolution takes place during the short irradiation time, however this limitation can be easily removed. 

As a result of radiation induced damage, cells can arrest the normal duplication cycle in any of the four phases, because some cycle checkpoint is not completed. We construct for this event a probability:
\begin{equation}
P_{arr} = \begin{cases}  1 & \text{if } z_1 > 0 \\ \\
                                  1-S     & \text{if } z_1  = 0  \end{cases}
\end{equation}

\noindent meaning that any cell having undergone DSB damage, or any cell entering senescence ($S>0$, see Eq. (\ref{equation:sene})) can be arrested.

Once in this arrested condition, a cell can take at each time step one of three alternatives (Fig. \ref{fig:flux}). Firstly, it can die and be eliminated by apoptosis: for example, it is known that the \emph{p53} protein, usually a regulatory transcription factor, can also trigger the expression of genes inducing apoptosis, when excessive DNA damage accumulates \citep{roos}. For this event, we introduce a probability:
\begin{equation}
P_{death} = \min \left[1,\left( \frac{z_1}{N_{crit}} \right)^{\alpha} \right]
\label{equation:death}
\end{equation}

\noindent with $\alpha>1$ a 'retarding' parameter, and $N_{crit}$ a critical threshold of accumulated DSBs leading to apoptosis ('mortality' parameter). The value of the threshold depends on the cell type and on the irradiation conditions, typical values being in the range $N_{crit} \sim 5 \div 50$. 

Alternatively the cell can restart its life cycle, with probability:
\begin{equation}
P_{restart} = e^{-\beta z_1}
\end{equation}

\noindent increasing at a rate $1/\beta \sim 0.5$ to 2, expressing the efficiency of the repair action, and getting equal to 1 when $z_1$ is back to 0. Or, finally, the cell can remain in the quiescent state with probability $(1-P_{death} - P_{restart})$ until the next time step. 

However, it is known that the reentry in the life cycle of a damaged cell can lead to a potentially cancerous cell, especially at lower levels of damage and attack to the tumour suppressor genes, events not immediately leading to a tumoral cell, but rather inducing a genomic instability, which is only one of the required components likely to induce a cancer. 
Such cells may be indicated as 'neoplastic'. 

\subsection{Calibration by irradiation experiments on human fibroblasts}

Normal human dermal fibroblasts (NHDFs) used in this study were PromoCell F-1MC from a 1-year-old Caucasian male. Cells were cultured in incubator at 5\% CO$_2$ and 37 $^{\circ}$C, in basal medium (FBM - Fibroblast cell Basal Medium, Lonza) with  2\% FBS (Foetal Bovine Serum), plus human fibroblast growth factor (hFGF), insulin at 5 mg/ml, and antibiotics Gentamicin 50 $\mu$g/ml and Amphotericin B 50 $\mu$g/ml. Cells were plated at 200,000 cells per 100-mm Petri dish, and cultures were always split at 70\% confluence with the Reagent PackTM by Clonetics (Hepes buffered saline solution, Trypsine/EDTA 25 mg and 10 mg per 100ml, TNS Trypsin Neutralizing Solution). Population doubling was calculated at each passage, after cell counting with a Thoma or Malassez counting chamber. The number of population doublings, $PD$, was calculated as: 
\begin{equation}
	PD = \frac{\ln(N_{coll}/N_{plat})}{\ln 2}
\end{equation}

\noindent N$_{coll}$, N$_{plat}$ being the number of collected and plated cells, respectively. 

Irradiations took place at the Varian Primus CLINAC of the "Oscar Lambret" center.  Cells were cultured either in 96-well plate, deposited on 4 cm of a plastic plate (equivalent tissue) to ensure electronic equilibrium. (Note that for a 20 MV accelerating tension, the photon spectrum is peaked at a much lower energy of about 3 MeV, with an average energy of about 10 MeV.) The photon beam was directed from below the table, with intensity adjusted to provide 200 monitor units at the isocenter of the culture dish.

After the irradiation, cells were fixed with formalin, and permeabilised with triton 1\% in PBS (Phosphate salt buffer: 3.2mM Na$_2$HPO$_4$, 0.5mM KH$_2$PO$_4$, 1.3mM KCl, 135mM NaCl, at pH=7.4). After washing, non-specific sites were blocked for 1h in 5\% skim milk diluted in PBS. Cells were then incubated in a solution of primary antibodies diluted at 1/100 or 1/200 (respectively for anti-53BP1 and anti-XRCC1) in the buffer (anti-53BP1 from Santa Cruz Biotechnology, SC-22760 and anti-XRCC1 from Santa Cruz Biotechnology, SC-11429). After several washes in PBS, cover slips were incubated in a solution of secondary antibodies A21206 anti-IG (rabbit) from Life Technologies, diluted at 1/500 and coupled to a fluorochrome (ALEXA FLUOR 488). Subsequently, cell nuclei were coloured by a solution of Hoechst 33342 (Sigma) 5 mg/l in PBS. 96-well plate were finally analysed by HCS Operetta (Perkin Elmer). Counting of XRCC1 and 53BP1 foci was carried out with an image analysis software (Columbus by Zeiss).


\section{Test cases in modelling cell culture irradiation}
\label{sect3}
In this section we  present two applications of our model, exploring only some of the wide range of possible simulations of cell evolution that it enables. We stress once more that such test cases are not meant to explore the complexity of the corresponding biophysical problems, \color{black} but they only serve as the first, necessary test phase of the model. \color{black}

\begin{figure}[t]
\begin{center}
\includegraphics[scale=0.45]{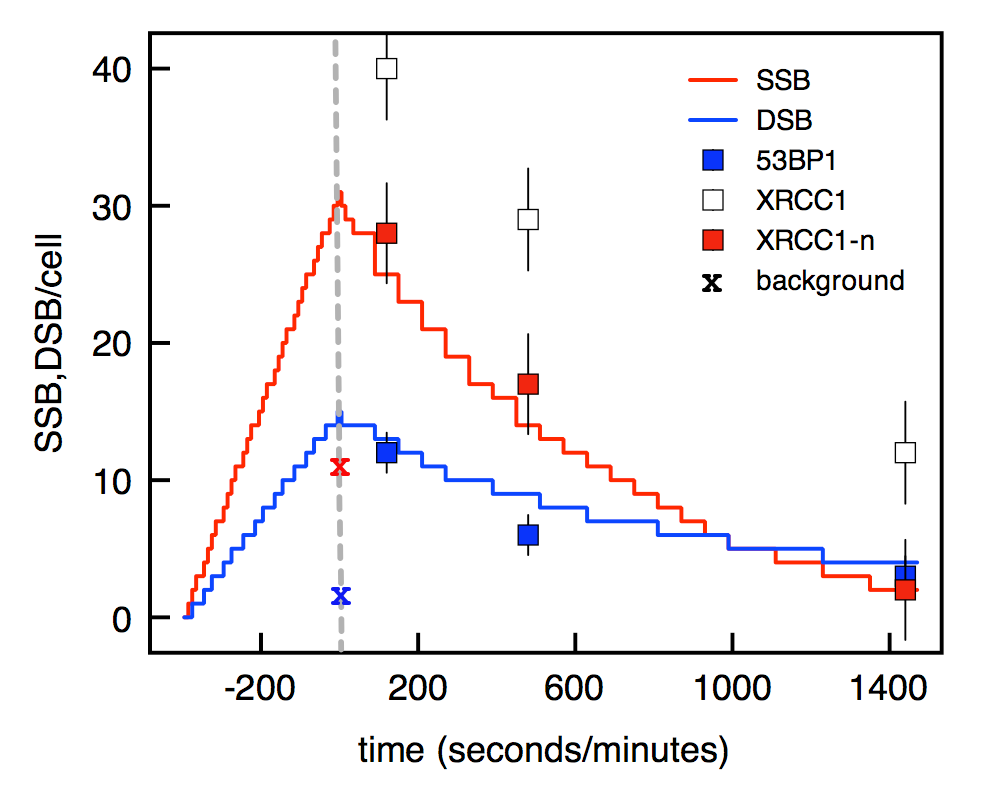}
\caption{Kinetics of the experimentally observed 53BP1 and XRCC1 foci after 2 Gy irradiation (symbols), with repair probabilities from Eq. \ref{equation:rpha} (histograms) adjusted to match the experimental kinetics. Crosses: background signal in absence of radiation (red=SSB, blue=DSB). White squares: XRCC1 raw data; red: XRCC1-n after subtracting the background at $t$=0; blue: 53BP1 data. Note the larger error bars for SSBs, pointing to a larger dispersion of the cell population for this type of lesions. The blue and red histograms correspond to the simulated model kinetics, with damage probabilities $\{p_1,p_2\}$, and repair probabilities $\{r_1,r_2\}$ adjusted to reproduce experimental irradiation data. The x-axis gives the time $t$ in seconds, for $t<0$, and in minutes, for $t>0$.}
\label{fig:expt}
\end{center}
\end{figure}

To begin with, we need to firstly adjust the model kinetics on our experimental results of cell irradiation. Figure \ref{fig:expt} shows with square symbols the results of the counting of DSBs (blue) and SSBs (red) after a nominal 2 Gy irradiation. Note that the experimental control value (indicated by crosses at $t$=0) is practically 0 for DSBs, while it is around 10 SSB/cell; such a background corresponds to daily generation of damage from sources other than the irradiation. After subtracting the background, it is seen that over the 24 hours about 25\% of the DSBs are still unrepaired (blue squares), while practically all the SSB have been repaired (red squares; white squares, before subtraction of control). However, it is worth noting that for SSB there remains a disparity between cells, as shown by the larger standard deviation in the histogram, some cells still showing $\sim$25 XRCC1 foci after 24h. The average value of about 6 DSB/Gy observed here is quite smaller than the accepted values, in the range of 15-35 DSB/Gy \citep{lobrich,schultz,awadi}. The source of such a discrepancy is not yet clear, it might be related to the particular kind of signalisation protein chosen for damage identification.

\color{black} For the computer simulation we will use a small test sample, made of $N_c$=500 cells distributed on a square lattice, and subject to a homogeneously distributed radiation dose. From our preliminary tests, such a size of population is already sufficient to obtain a good statistics on the system response. \color{black} In the same Fig. \ref{fig:expt} we show two calculated histograms for creation ($t<0$) and reparation ($t>0$) of DSB (blue) and SSB (red). We calibrated the probabilities $p_1$ and $p_2$ of generating a DSB or a SSB, respectively, as well as the corresponding repair probabilities $r_1$ and $r_2$, by using the experimental data as a reference. For the sake of simplicity, we assume the dose of 2 Gy, at a rate of 3 Gy/min (lower range of a typical LINAC machine), to be delivered by a monochromatic spectrum of gamma-rays at 20 MV. (Note that the x-axis in the figure give the time in seconds, for $t<0$, and in minutes for $t>0$). The best reproduction of the experimental data is obtained with $p_1=3.0 \times 10^{-4}, p_2=6.5 \times 10^{-4}, r_1=1.5 \times 10^{-3}, r_2=2.5 \times 10^{-3}$. \color{black} The resulting histograms have a nearly exponential decay (as expected from Eq.(\ref{equation:exprob})), with the respective $t$=0 values giving the $p_{\nu}$, and the decay constants giving the $r_{\nu}$. Also, note that since we have 3+3 experimental data points, there is no over fitting of the four parameters. \color{black}

\begin{figure}[t]
\begin{center}
\includegraphics[scale=0.35]{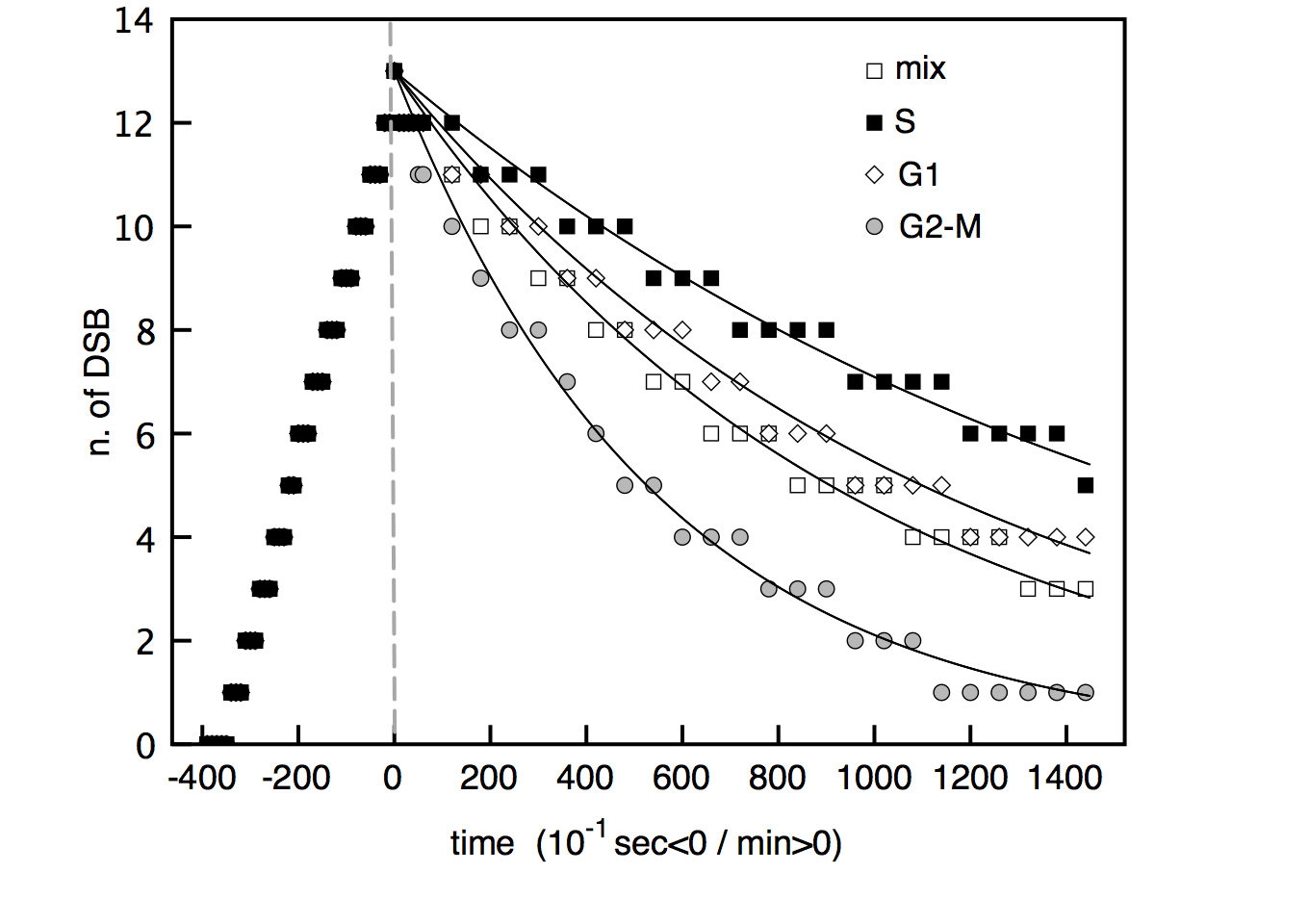}
\caption{Effect of including a cell-phase-dependent repair probability (see Eq.(\ref{equation:rpha})). 'Mix' is the average, with repair functioning in all cell phases, 'S', 'G1', 'G2-M' refer to a repair probability restricted only to the corresponding cell phase. Continuous lines represent exponential fits to each set of symbols.}
\label{fig:scale}
\end{center}
\end{figure}

Figure \ref{fig:scale} shows the effect of using a repair probability that depends on the cell phase. The open squares for the 'mix' case, in which the repair probability follows Eq. (\ref{equation:rpha}), correspond to the best fit already given in the previous Fig. \ref{fig:expt}. For the other cases, it can be seen that by restricting the repair to S-phase only a slower recovery of damage is obtained, while, at the other extreme, the faster recovery would correspond to concentrating the repair only in the G2 and M phases. \color{black} In the remainder of this work, we will use the 'mix' option corresponding to the best fit. \color{black}

After calibrating in this way the kinetics of our model, we will present a first example in which we apply the model to reproduce the well-known results of typical cell survival curves, following irradiation with increasing levels of dose. 
Then, in a second example we will test the ability of the model to incorporate spatial information, by developing a set of simulations of the so-called "bystander" effect, namely the alteration in the response of non-irradiated cells when some nearby cells are irradiated.

\subsection{Survival curves for irradiated cells}

The "linear-quadratic" (LQ) model is the most widely accepted mechanistic model of cell killing by radiation (see e.g. \citet{dale}). \color{black} While being a strictly phenomenological representation of experimental data about cell survival as a function of the total dose, the LQ is often justified by the concept of binary DSB misrepair: \color{black} since it is believed that the main chromosome aberrations ("dicentric") should likely occur after the wrong rejoining of a pair of close-by DSBs \citep{wlodek}, it may occur that prolonged exposure to radiation would allow only one of the DSB to be repaired, before the second is generated. \color{black} Although this is the most usual way to provide a motivation to the standard LQ approach, \color{black} alternative biological explanations have also been advanced, such as the repair-misrepair \citep{tobias}, the lethal-potentially lethal \citep{curtis}, or the two-lesion kinetic model \citep{stewart}, which give practically the same numerical results of the LQ; as well as alternative models, variously based on concepts of saturation of the repair capacity \citep{sanchez,cucinotta}. Overall, it is generally acknowledged that the biophysical basis of all such models rests on quite speculative assumptions about the microscopic events leading to the observed cell response. Their most important value, notably for the LQ model, rather resides in their empirical utility in dose treatment planning (for a comparison among the various models, see e.g. \citet{brenner}). \color{black} However, one cannot say to have learned much about the biophysics of DNA damage from empirical models of this kind.\color{black}

\begin{figure}[t]
\begin{center}
\includegraphics[scale=0.32]{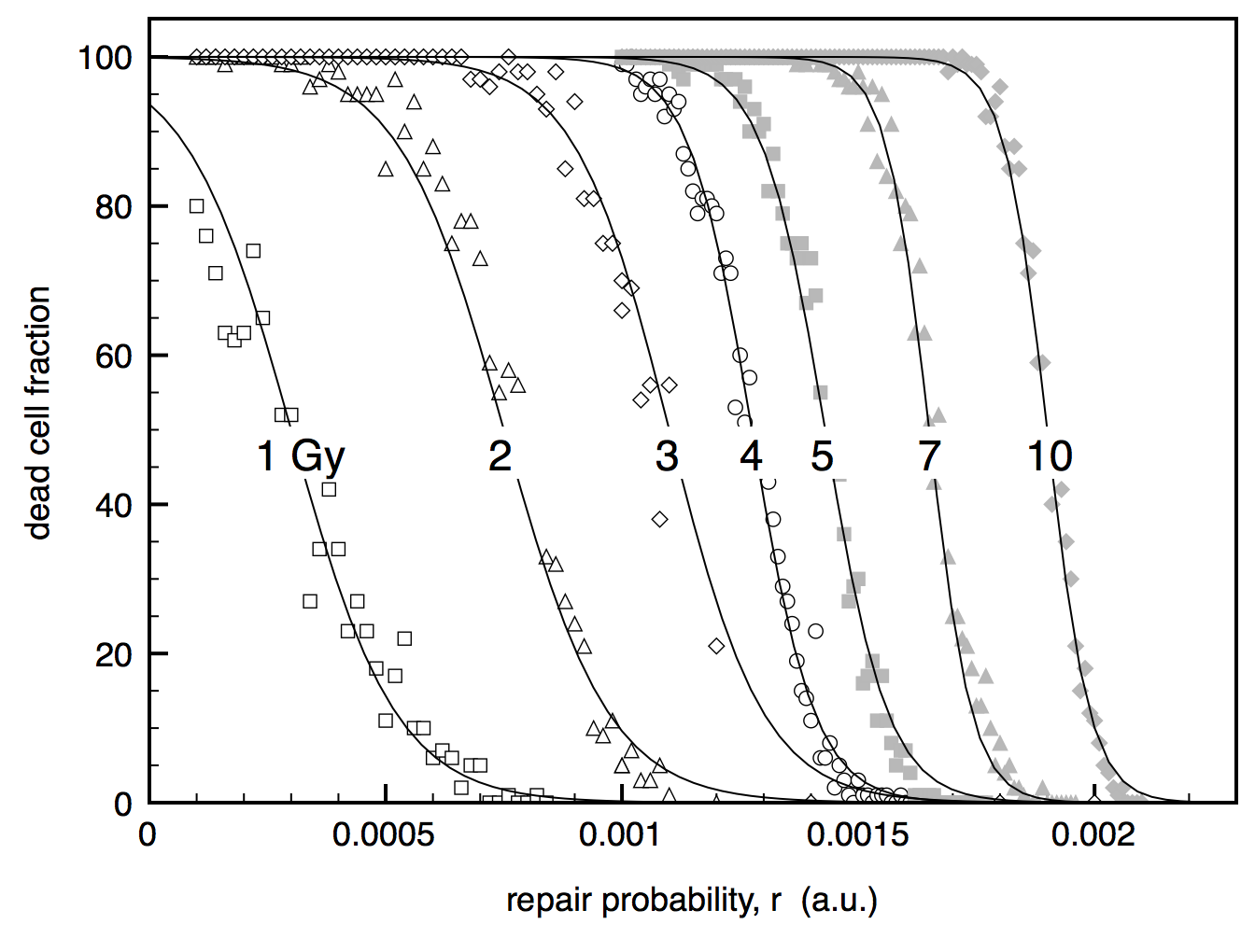}
\caption{Plot of the cell death fraction from a colony of $N$=500 cells, as a function of the DSB repair probability $r_1$, for simulated irradiations at dose levels between 1 and 10 Gy (indicated for each set of data), at a rate of 3 Gy/min. Damage probabilities $p_1,p_2$, and repair probability $r_2$, are fixed at the same values reproducing experimental irradiation data in Fig. \ref{fig:expt}. Cells are observed for 24h after the irradiation, and the mortality threshold is fixed at $N_{crit}$=5 (see Eq. (\ref{equation:death})). Continuous curves are best fits with a sigmoidal function, for each set of simulation data.}
\label{fig:multicurv}
\end{center}
\end{figure}

We used our simulation model to attempt an interpretation of typical cell survival curves, however without including any  of the above detailed (but speculative) microscopic assumptions about the typology and evolution of DSB generation. As an example, we will test the hypothesis \citep{sanchez,cucinotta} that the repair probability may be dependent on the dose, by means of some 'saturation' mechanism, which could be further traced back to a molecular origin. However, in our simulations we will not explicitly introduce the saturation hypothesis. Rather, we will fix some damage threshold $N_{crit}$, and explore the cell survival fraction as a function of the radiation dose \textit{and} the DSB repair probability $r_1$. The other ingredients are those already given in Sect. 2, namely DSBs are taken to occur independently, with Poisson-like statistics, followed by  a simply exponential repair capability. No further fitting is needed, since we use the values of probability $\{p_1,p_2,r_2\}$ corresponding to the previous fitting (Sect. 3.1 above) of our own experimental data at the dose of 2 Gy. 
 In this way, we aim at showing (i) that the model is data-independent (since we use data from a particular fitting, to reproduce  data obtained in entirely different experiments), and most importantly (ii) that the model can autonomously generate a non-predefined response, \color{black} thus providing a biophysical interpretation to the experimental data.\color{black}

Figure \ref{fig:multicurv} shows the evolution of the cell death fraction as a function of the 
value of $r_1$, for irradiations at different dose levels $\Delta$=1,2,...10 Gy, at the rate of 3 Gy/min. The mortality threshold is fixed at a quite low value, $N_{crit}$=5, meaning that on average 5 DSBs in a cell are already enough to lead to apoptosis. Cells are observed for a time of 24h after the irradiation. For the sake of simplicity, we are using a phase-independent repair rate, i.e.  the 'mix' in Fig. \ref{fig:scale}. All the curves for the various simulations display the same behavior, namely the death probability is close to 1 for low values of $r_1$, and it rapidly decreases around a critical value of $r_1$, depending on the dose. Upon increasing $r_1$ all the curves end up to zero mortality, meaning that the repair probability gets large enough to provide a full healing of DSB damage within the 24h time window. Note that for the lowest dose simulated of 1 Gy, the plot does not start at 100, since the number of DSBs generated by such a small dose is below $N_{crit}$=5.


\begin{figure}[t]
\begin{center}
\includegraphics[scale=0.37]{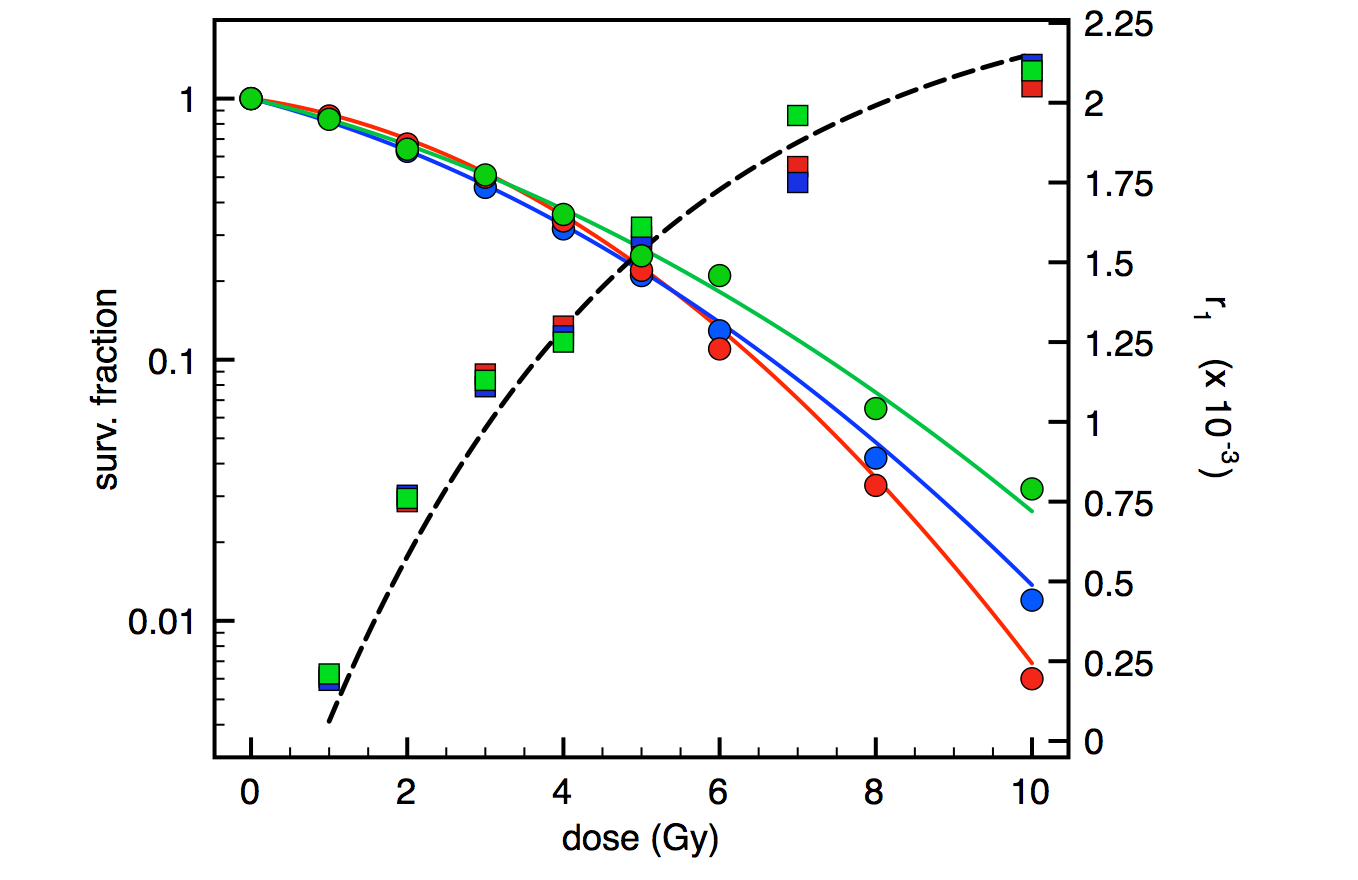}
\caption{Left y-axis: plot of the experimental data of survival fraction for V79 (blue dots) and FaDu (red dots) cells, from \citet{brita}, and for X1 cells (green dots) from \citet{mama}. Continuous curves represent a fit to our simulation data. Right y-axis: square symbols: values of $r_1$ giving the best fits (color curves) to the experimental survival fraction, for each dose value; symbol colors correspond to each experimental and fitted curve. The dashed line is an exponential fit to the $r_1$ values from the model.} 
\label{fig:doublepl}
\end{center}
\end{figure}

Now, 
we take some typical cell survival curves from published irradiation experiments, and compare the experimental data to our simulated survival fractions from Fig. \ref{fig:multicurv}. Namely, for each value of dose $\Delta$, we extract from the corresponding curve in Fig. \ref{fig:multicurv} the value of $r_1$ which allows to best fit the experimental survival fraction at that dose. Figure \ref{fig:doublepl} shows three such experimental curves. Two are from the work of \citet{brita}, obtained by irradiation of V79 (blue) and FaDu (red) chinese hamster cells with a LINAC, at dose rates of 5.01, 9.99 and 29.91 Gy/min (see left y-axis). The three colour curves represent our model fit to the experimental data, as proposed in the original works. In such experiments, no dependence of the survival fraction on the dose rate was evidenced, in fact the LQ fits for the three different rates are practically identical. Note that such dose rates are much larger than those used in classical survival curve experiments by smaller radioactive sources: as an example we include in the Figure also data by \citet{mama} (green) on X1 chinese hamster cells irradiated by a 250 kV x-ray tube. 

In the same Figure \ref{fig:doublepl}, right axis, we also plot the values of $r_1$ which provide the closest fit to each curve (blue squares for the V79, red squares for the FaDu, green squares for the X1), at each value of dose. It can be seen that the best values of $r_1$ are not at all constant, but follow a clearly saturating pattern, i.e. increasing steadily at low doses and aiming at a constant value (about $r_1 \sim  2.4 \times 10^{-3}$) at higher doses. The dashed curve in the figure is an exponential fit to these $r_1$ values, of the type $a-b \exp(-c r_1)$, with $a$=2.4, $b$=3 and $c$=0.25. 

The important point here is that our dynamical simulation model is capable of producing survival curves in \color{black} very good \color{black} agreement with various sets of experimental data, without making \textit{a priori} assumptions \color{black} about the biophysical mechanisms underlying. In a sense, we let the cells 'choose' their behavior, and we look which particular behavior corresponds to the observed experimental data. In this way, \color{black} the model \textit{suggests} a biophysical mechanism underlying the observed shape of the survival curves: rather than by making hypotheses of, e.g., linear-plus-quadratic components (of doubtful physical origin), the model \color{black} only includes biologically observable damage (SSBs and DSBs, although only the latter are considered as lethal, according to the current knowledge). With this initial hypothesis, the model is able to \color{black} suggest that cells may adopt a molecular repair response that is exponentially saturable. One could then further speculate about the molecular connection, for example imagining that there is a finite supply of repair proteins in the cell nucleus, which fails when the damage is too extended. \color{black} While we used here this set of computer simulations only as an example of application of the model, a more complete study of this effect is underway, and we defer further considerations to a future work.\color{black}

\subsection{Computational model for 'bystander' effect in radiobiology : long-range vs. short-range diffusion.}


\begin{figure*}[t]
\begin{center}
\includegraphics[scale=0.29]{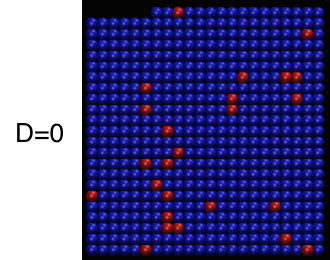}
\includegraphics[scale=0.1]{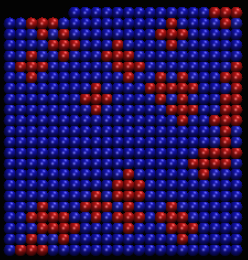}
\includegraphics[scale=0.1]{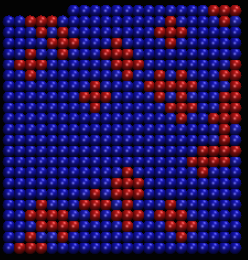}
\includegraphics[scale=0.1]{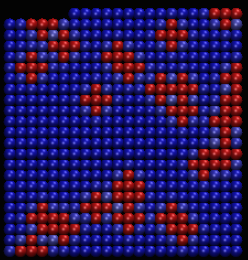} \\
\includegraphics[scale=0.29]{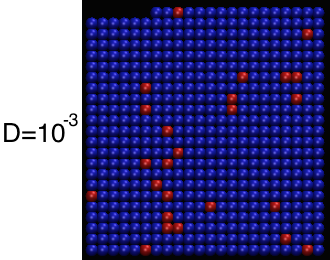}
\includegraphics[scale=0.1]{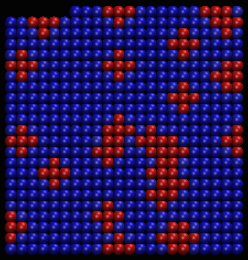}
\includegraphics[scale=0.1]{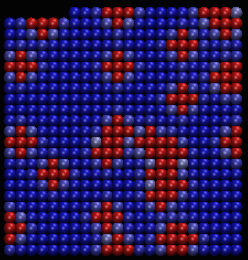}
\includegraphics[scale=0.1]{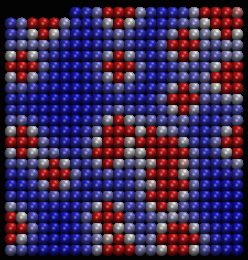} \\
\includegraphics[scale=0.29]{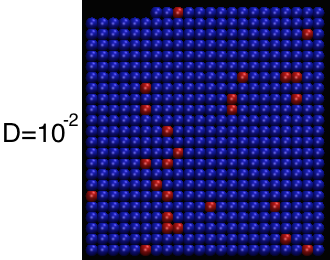}
\includegraphics[scale=0.1]{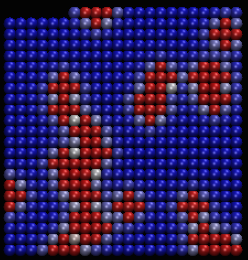}
\includegraphics[scale=0.1]{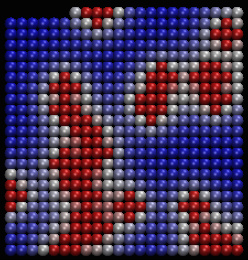}
\includegraphics[scale=0.1]{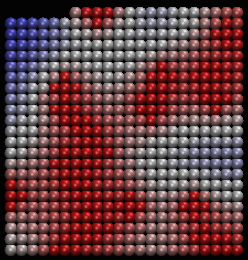} \\
\includegraphics[scale=0.29]{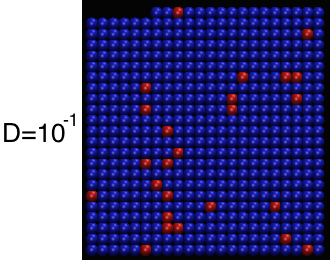}
\includegraphics[scale=0.1]{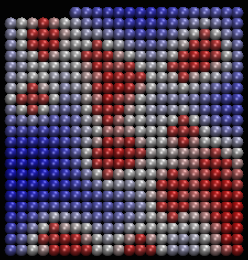}
\includegraphics[scale=0.1]{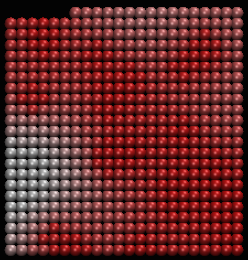}
\includegraphics[scale=0.1]{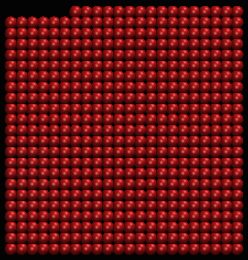}
\includegraphics[scale=0.45]{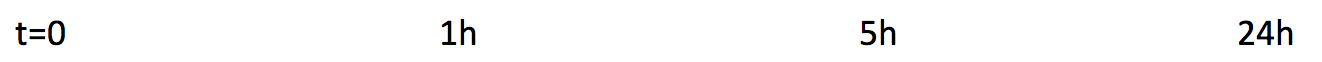}
\caption{Concentration maps from the simulation of bystander effect in a colony of 500 cells with a randomly distributed 5\% fraction of irradiated cells. Rows from top to bottom: diffusion coefficient for non-irradiated cells $D^B$=$\sim$0, 0.001, 0.01, 0.1. The top row represents a \emph{local} bystander effect, requiring direct cell-cell contact; the lower rows correspond to the \emph{global} effect, with increasing extracellular diffusion velocities. Columns from left to right: simulation time $t$=0, 1, 5, 24h. The continuous color coding represents the concentration of the secreted factor in each cell (blue $c^B$=0, white $c^B$=0.5,red $c^B$=1).} 
\label{fig:celmap}
\end{center}
\end{figure*}

The so-called "bystander" action is seen to occur in experiments performed under widely different conditions, both \emph{in vitro} ad \emph{in vivo}, and it appears to fall under two broad categories: (i) \emph{local}  action over short cell-cell distances; (ii) \emph{global} action over longer distances.


The cellular effects of such bystander actions generally involve an upregulation of the metabolism of co-affected cells, such as the oxidative pathways \citep{narayan,mother}, the p53 damage-response pathway \citep{azzam}, increased levels of interleukin-8 \citep{naraya2}, AP endonuclease \citep{iyer1}, TGF $\beta$1 \citep{iyer2}, and others. Notably, increase in cell proliferation activity has often been associated with such increased metabolism \citep{iyer2,iyer1,gerash1,gerash2}. This latter finding motivates a second application of our model, in order to show the importance of coupling the damage-repair capability with diffusion and cell topology.

We performed a series of coupled irradiation-diffusion simulations on a sample colony of $N_c$=500 cells, with the same probability parameters as above. To simulate co-culture of irradiated and non-irradiated cells, randomly picked sub-populations corresponding to 5, 10, 20, 50, 75 and 90\% of the colony were irradiated with the same protocol of Sect. 3.1. The bystander effect was simulated in the two variants above described, namely  (i) local and (ii) global, by allowing the irradiated cells to diffuse a generic pro-mitogenic factor (such as TGF $\beta$1), here labelled B for 'bystander', whose concentration $c^B$ is supposed to increase over time in each cell, according to the diffusion equation (\ref{equation:fick}). The source term was set to $s^B=1$ in irradiated cells, and $s^B=0$ elsewhere. The increase in proliferation was linked to the concentration, by shifting each local cell clock as:
\begin{equation}
t_{d'} = t_d + \Delta t (1 + \gamma c^B)
\label{equation:tshift}
\end{equation}

\noindent with $\gamma$ an empirical efficiency factor, ranging from 0.1 to 2. In this way, cells that accumulate larger concentrations of the 'B' factor may anticipate their  time to the next division, up to twice the normal rate. In the same spirit, we also tested the effect of a cytotoxic factor (e.g., reactive oxygen species, hydrogen peroxide, nitric oxyde) leading to proliferation arrest, by setting $\gamma$=-0.5, -1. Moreover, we tested the possibility that the 'B' factor may also induce apoptosis when exceeding some threshold (for example, $c^B \geq 0.9$), by relating the concentration to $P_{death}$ in Eq.(\ref{equation:death}) as:
\begin{equation}
 P_{death} = \frac{1}{1+e^{-100(c^B-0.9)}}
\end{equation}

Then, the 'local' variant of the bystander effect (i), was represented by assigning diffusion coefficients $D^B$=1 and $D^B\simeq$0 (in arbitrary units) to the irradiated and non-irradiated cells, respectively. The 'global' bystander effect (ii), was instead represented by assigning $D^B$=1 to irradiated, and increasing values of $D^B$=10$^{-3}$, 10$^{-2}$, or 10$^{-1}$ to the surrounding, non-irradiated cells. An example of the typical effect of such choices is represented in the 2D cell maps of Figure \ref{fig:celmap}.

\begin{figure}[t]
\begin{center}
\includegraphics[scale=0.2]{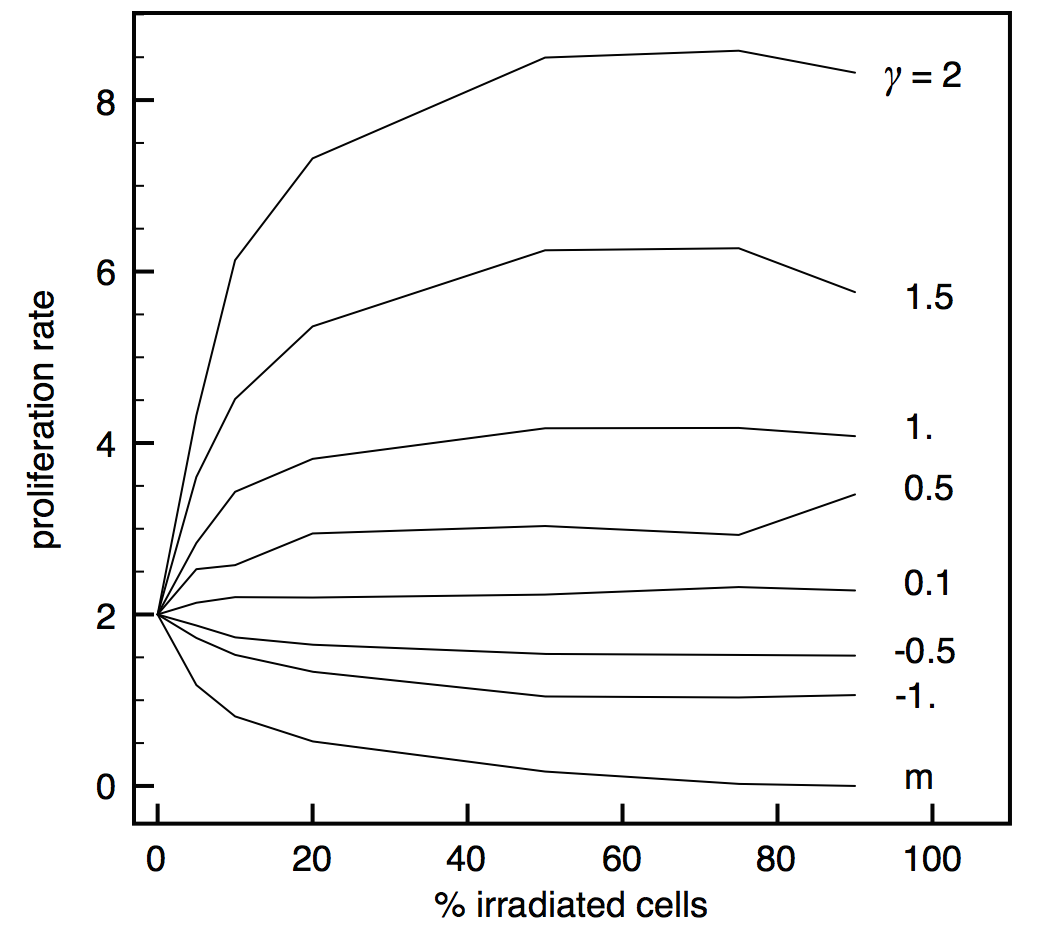}
\includegraphics[scale=0.2]{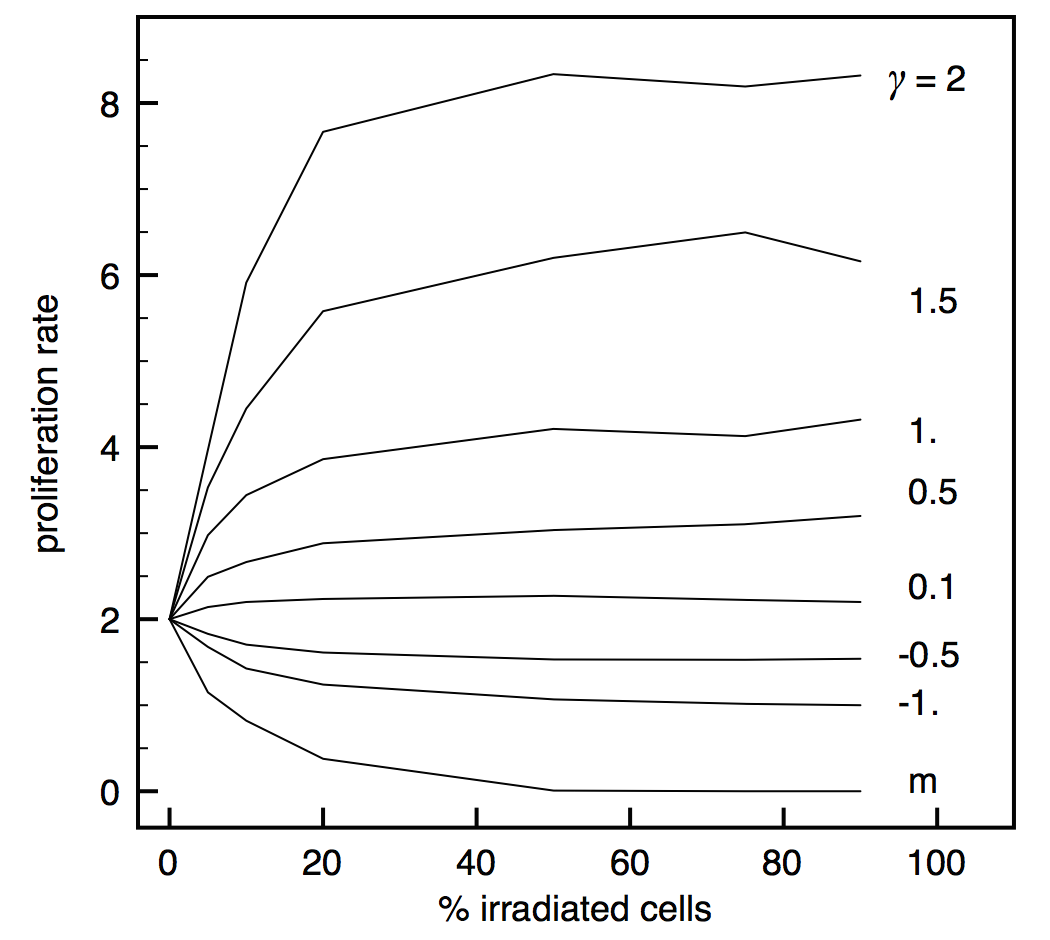} \\
\includegraphics[scale=0.2]{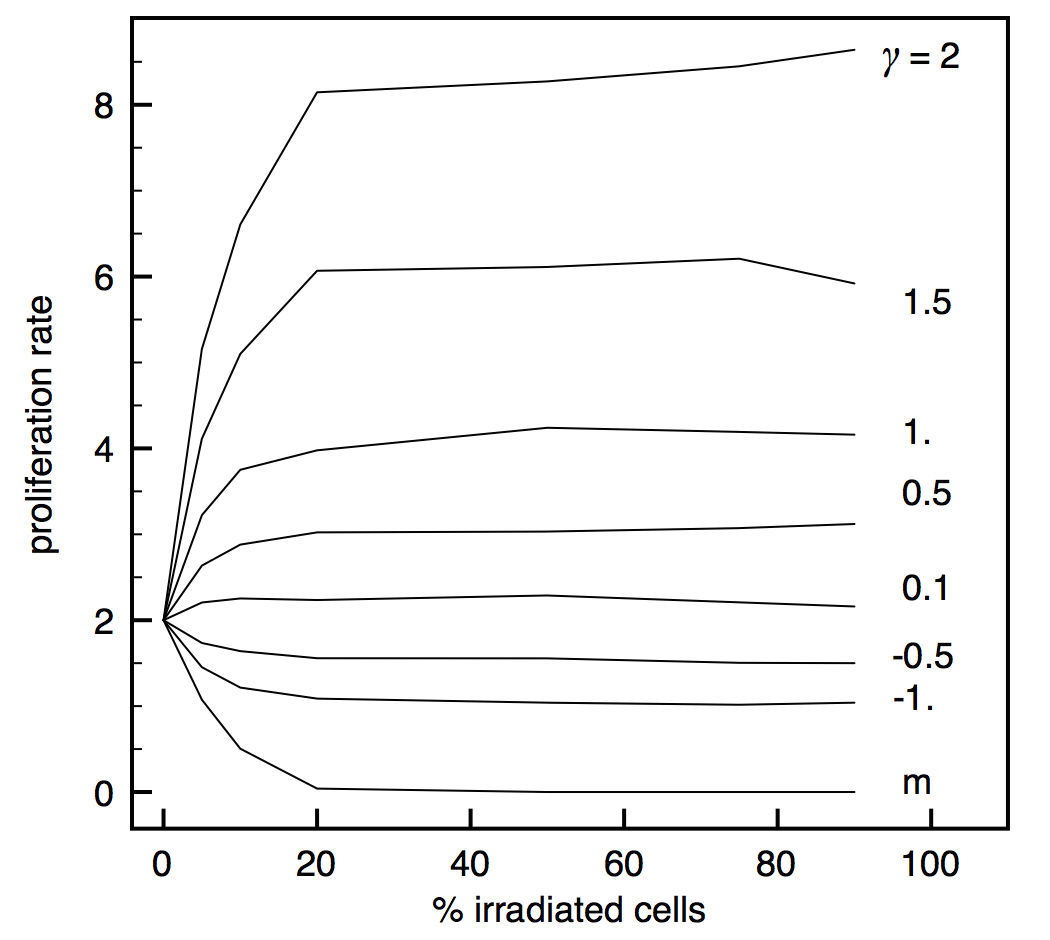}
\includegraphics[scale=0.2]{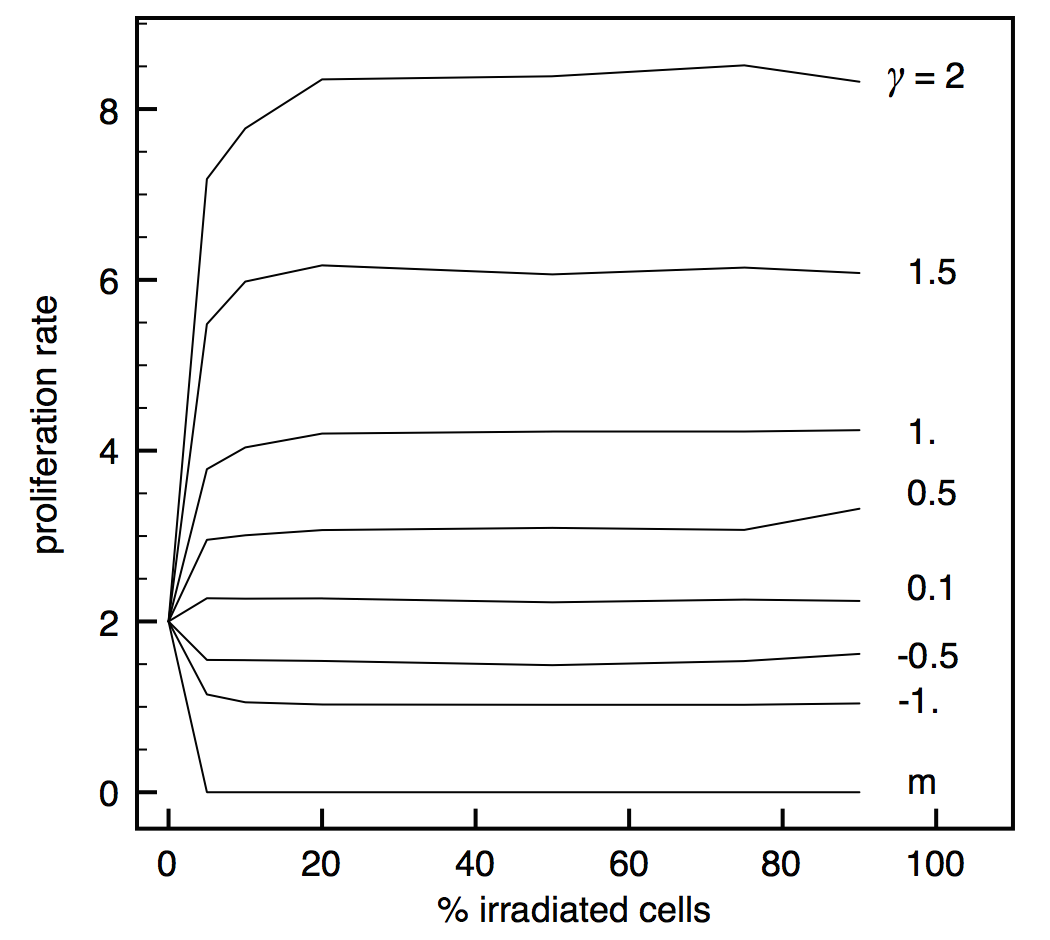} 
\caption{Plot of the duplication rate of non-irradiated cells after 24h, in a colony of 500 cells including fractions of irradiated cells ranging from 5 to 90\%. In each simulation, the diffusion coefficient of irradiated cells is always $D^B$=1. The diffusion coefficient of non-irradiated cells varies in each panel. Top row, left $D^B \sim$0, right $D^B$=0.001; bottom row, left $D^B$=0.01, right $D^B$=0.1. In each panel, the various curves correspond to different values of $\gamma$ in Eq.(\ref{equation:tshift}).} 
\label{fig:byst1}
\end{center}
\end{figure}

\begin{figure}[t]
\begin{center}
\includegraphics[scale=0.3]{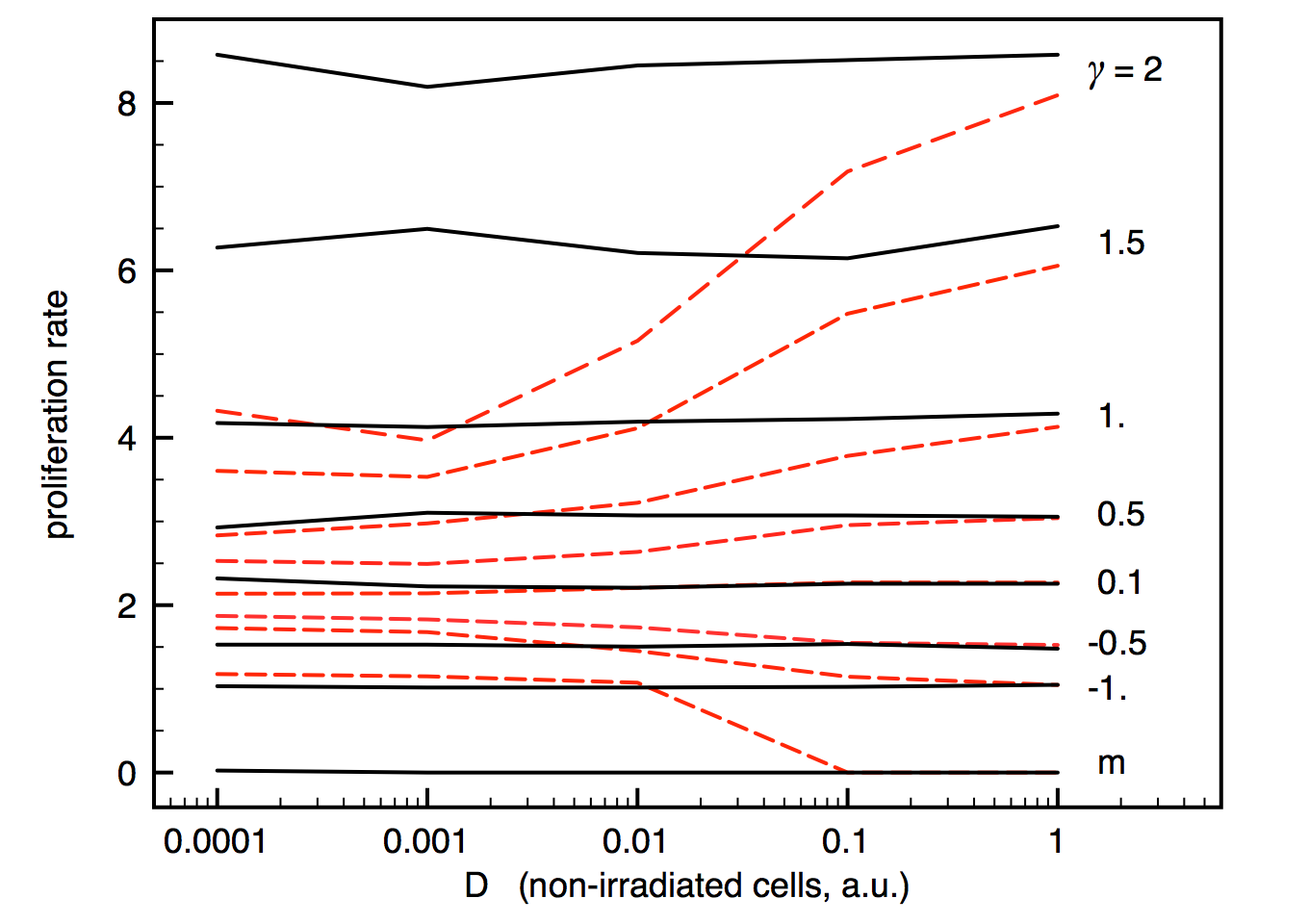}
\caption{Plot of the duplication rate after 24h, in a colony of 500 cells including 5\% (red, dashed) or 75\% (black, full curves) fraction of irradiated cells, as a function of the diffusion coefficient of non-irradiated cells. The diffusion coefficient of irradiated cells is always $D^B$=1. The various curves correspond to different values of $\gamma$ in Eq.(\ref{equation:tshift}).} 
\label{fig:byst2}
\end{center}
\end{figure}

The results of the simulations are summarized in Figures \ref{fig:byst1} and \ref{fig:byst2}. The first one shows the duplication rate of non-irradiated cells after 24h, in a colony in which variable proportions of irradiated cells (in abscissa) are randomly mixed. The four panels correspond to the four values of $D^B \simeq$0, =0.001, 0.01, 0.1. The curves in each panel correspond to different values of $\gamma$, the coefficient relating the concentration of the B factor to the proliferation rate. From such data, only minor differences between the 'local' and 'global' bystander effect could be inferred: all the families of curves resemble each other, all showing a sizeable increase in the duplication rate as a function of $\gamma$ (the normal proliferation rate in the absence of irradiation would be 2 after 24h). For $\gamma$ approaching zero, the proliferation rate goes back to constant, around its normal value of 2. Negative values of $\gamma$ give instead reduced proliferation rates, with the cell population being reduced to some constant value, since part of the cells go into the arrested condition. The case in which the B factor induces apoptosis is the curve labelled 'm' (for 'mortality') in each panel. It can be seen that in this case the cell colony can be completely killed after 24h. In this latter case, some correlation between the fraction of irradiated cells and the diffusion coefficient exists: at the smaller values of $D^B$, the mortality rate is nearly exponential with the fraction of irradiated cells, while at larger diffusion coefficients the mortality grows very quickly, and already a fraction of 10 to 20\% of irradiated cells is sufficient to propagate very effectively the simulated cytotoxic factor. However, by comparing the upper-left and the lower-right panel, at least a qualitative difference can be appreciated, the local effect showing a dependence of the proliferation rate on the fraction of irradiated cells, which is practically absent in the global effect at the highest values of $D^B$.

One has to look at Figure \ref{fig:byst2} to obtain a much clearer evidence of a sizeable difference between the local and global versions of the bystander effect. The plot represents two families of curves, each one corresponding to different values of $\gamma$ as in the previous Fig. \ref{fig:byst1}. However, in this case the abscissa represents the values of $D^B$ in logarithmic scale. The two families of curves are for two extreme values of fractions of irradiated cells in the colony, a small 5\% (red lines, dashed) and a rather large 75\% (black lines, full). It can be clearly appreciated that, for a high fraction of irradiated cells, both the local and global effect give results that are practically independent on the value of $D^B$, the main parameter to determine the proliferation rate being only the value of $\gamma$. However, at the smallest fractions, and for the larger values of $\gamma >$1, there is a marked dependence of the proliferation on $D^B$, the 'global' bystander effect leading to much higher proliferation rate than the 'local', indeed more than doubled for $\gamma$=2. This dependence seems less apparent for the negative values of $\gamma$.

\color{black} Such a result of the model is not at all surprising, and could have been readily predicted simply on the basis of logical reasoning: if the number of 'source' (i.e., irradiated) cells is large, there is practically no difference between a local and a global action; if the number of source cells is quite sparse, instead, a global action is necessarily more effective than a local one, for a given amount of time. Being only a test problem, this is an indirect proof that the model works correctly. \color{black} Clearly, our naive, two-parameter representation of the bystander effect is far from capturing the complexity of this elusive phenomenon, which requires instead a much deeper investigation. We used it here, just for the sake of demonstrating the power of a theoretical model of cell evolution capable of coupling damage and repair information at the single-cell level, together with spatial and topological information at the multicellular level. 

\section{Discussion and Conclusions}
\label{sect4}

In this work we developed and tested an agent-based model of cell evolution under the action of cytotoxic treatments, aimed at including the major features of cell cycle and proliferation, cell damage and repair, chemical diffusion. In this first implementation of the model, the cells-agents live in a two-dimensional lattice, with a diamond-neighbour topology. Cell evolution is mathematically driven by a Markov chain, with cells taking on a series of discrete internal states from 'normal' to 'inactive' (or 'dead'). Probabilistic laws are introduced for each type of event a cell can undergo during its life cycle: duplication, arrest, apoptosis, senescence, damage, healing. The system is simulated with a time-forward explicit algorithm, with a Monte Carlo sampling of the various probability distributions.

We firstly calibrated some of the free parameters with a series of cell irradiation experiments, carried out in a clinical LINAC at 20 MV and quite high fluence ($\sim$3 Gy/min). Single- (SSB) and double-strand breaks (DSB) in irradiated cells were counted by means of automated image analysis, after staining with XRCC1 (for SSBs) and 53BP1 (for DSBs) proteins. The resulting damage and repair kinetics were deduced, and included in the model parametrization. 

The very first test of the model was directed at reobtaining the cell survival curves from a number of published low- and high-dose irradiation experiments. Such curves are generally interpreted on the basis of the so-called "linear-quadratic" model, which presumes the DSB damage to occur in two different forms, a single-hit and a double-hit mode (hence the two terms in the LQ model). We could obtain a very good representation of the same cell survival curves, however without assuming any special hypothesis about the types of DSBs or other. We only let the cell DSB-repair probability as a variable, and we obtained that in order to fit the experiments such a probability must saturate as a function of the increasing dose, as previously suggested e.g. by \citet{sanchez} and \citet{cucinotta}. This simple test already shows the power of the simulation method, in that it is capable of suggesting a biophysical behavior that cells may adopt (an exponentially-saturating DSBs repair capability), instead of relying on various pre-conceived hypotheses. 

As a second test, we attempted to simulate 
 the so-called 'bystander' effect in radiotherapy, namely the possibility that non-irradiated cells located in proximity of irradiated cells may develop a sequence of metabolic response similar to these latter. We tested the two extreme, opposing hypotheses of a 'local' bystander effect, activated by short-range intercellular diffusion, versus a 'global' effect, activated by the long-range diffusion, e.g. of some factor secreted from the irradiated cells in the extracellular fluid. Even if our simulation was exceedingly simple, and many important factors were explicitly neglected, we could demonstrate some sizeable difference in the proliferation rate of non-irradiated cells, the increase in proliferation being much larger for the global than for the local effect, at least for a relatively small initial fraction of irradiated cells in contact with normal cells.

In this first study, primarily devoted to introducing and testing our agent-based model, we did not use the information about the other type of lesions already included in the algorithm, namely the single-strand breaks, for which we anyway performed the experimental calibration in the same way as for the double-strand breaks. We are however planning to use such a feature in a forthcoming study of tumour dormancy, as well as the other features not yet fully developed, such as the cell motility, which would be essential for example in a context of tumour heterogeneity studies.


\vspace{1cm}

\subsubsection*{Acknowledgements} 

M.T. thanks the Region Nord/Pas-de-Calais and the President of the University of Lille for a doctoral grant. C.A. and F.C. thank the SIRIC OncoLille for the grants SENDORMIR (Emergents 2013) and MODCEL (Emergents 2014). Computer resources provided by CINES Montpellier, under contract c2015-077225. We gratefully thank T. Lacornerie (Centre "Oscar Lambret", Lille) for his kind assistance with irradiation manipulations, and N. Martin and J. Nassour (Institut de Biologie de Lille) for their kind assistance with biology experiments.

\vspace{1cm}

\newpage

\begin{table*}
\begin{footnotesize}
\begin{tabular}{llp{15mm}p{70mm}l}
\textbf{Symbol}&\textbf{Variable}&\textbf{Units}&\textbf{Role}&\textbf{Values}\\
\hline
$N$ & lattice size &  & linear size of the square lattice & --- \\
$i$ & site index &  & label of lattice site & [1,$N^2$] \\
$N_c$ & number of cells &  & running size of the cell population & --- \\
$u(i)$ & cell index &  & label of cell $u$ on a lattice site & [1,$N_c$] \\
$\Delta$ & dose &  & amount of energy supplied over a given time interval & $[0,\infty]$ \\ 
\textbf{n} & cell state vector & & vector containing all local cell parameters & --- \\
$t$ & global time & seconds & universal 'laboratory' time of the simulation & $[-\infty,\infty]$\\
$\Delta t$ & time step & seconds, minutes & discrete simulation time increment & [0.001-0.1]\\
$t_d$ & cell time & minutes & time clock local to each cell & [0,1440] \\
$t_0$ & duplication time & minutes & time since last duplication for each cell & $[0,\infty]$ \\
$\tau$ & duplication time constant & minutes & see Eq.(\ref{equation:pdupl})  & 30 \\
$\phi$ & cell phase index &  & defines the cell phase (G0,G1,S,G2,M) & [0,4] \\
$\lambda$ & cell state index & & normal, stem, cancer, arrested, dead, neoplastic & [1,6]\\
$\nu$ & damage type &  &  double- or single-strand breaks (DSB, SSB) & [1,2] \\
$z_{\nu}$ & damage counter &  & counts accumulated number of defects of type $\nu$ & $[0,\infty]$ \\ 
$p_{\nu}$ & damage probability &  & probability of generating a defect of type $\nu$ & $[0,1]$ \\ 
$r_{\nu}$ & repair probability &  & probability of healing a defect of type $\nu$ & $[0,1]$ \\ 
$N_{crit}$ & critical damage &  & number of lethal damage (DSBs) above which a cell undergoes apoptosis & [5,100] \\
$D$ & n. of duplications &   & number of duplications undergone by each cell & $[0,\infty]$ \\
$S$ & senescence factor &  & describe the senescence of each cell & [1,0] \\
$D_0$ & n. of duplications &   & n. of duplications at which senescence starts & 30 \\
$D_s$ & n. of duplications &   & n. of duplications at which senescence is complete & 60 \\
$P_n$ & cell state probability &  & probability that cell is in state \textbf{n} at a given time & [0,1] \\
$P_{dupl}$ & duplication probability &  & controls probability of duplication for each cell & [0,1] \\
$P_{arr}$ & arrest probability &  & controls probability of arresting a cell for DSB accumulation or senescence & [0,1] \\
$P_{death}$ & death probability &  & controls probability of apoptosis when DSBs in a cell approach $N_{crit}$ & [0,1] \\
$\alpha$ & retarding parameter &  & slows cell killing by DSB accumulation & $>1$ \\
$P_{restart}$ & restart probability &  & controls probability of restarting cell cycle from an arrested state & [0,1] \\
$\beta$ & accelerating parameter &  & accelerates cell repair capability to promote restarting & [0.5-2.] \\
$c^{\mu}_{u(i)}$ & chemical concentration &  & instantaneous concentration of species $\mu$ in cell $u(i)$ & $[0,\infty]$ \\
$s^{\mu}_{u(i)}$ & source concentration &  & constant source of species $\mu$ in cell $u(i)$ & $[0,\infty]$ \\
$\theta^{\mu}$ & diffusion time & minutes & inverse diffusion coefficient for species $\mu$ & $[0,\infty]$ \\
$c^B$ & B-factor concentration &  & instantaneous concentration of 'bystander' pro-mitogenic factor & $[0,\infty]$ \\
$s^B$ & B-factor source &  & constant source of 'bystander' pro-mitogenic factor & $[0,1]$ \\
$D^B$ & B-factor diffusion coefficient &  & reciprocal of $\theta^B$, the diffusion time of the 'bystander' pro-mitogenic factor across the cell membrane & $[0,1]$ \\
$\gamma$ & coupling parameter &  & couples concentration of B-factor to cell duplication time & [0.1-2.] \\
\hline
\end{tabular}
\end{footnotesize}
  \caption{List of the principal model variables. Unless differently indicated, all the variables are adimensional.}
\end{table*}


\begin{thebibliography}{00}



\bibitem[Aguirre-Ghiso(2007)]{aguirre}{Aguirre-Ghiso, J. A.: Models, mechanisms and clinical evidence for cancer dormancy. Nature Rev. Cancer 7, 834-846 (2007)}

\bibitem[Albright(1989)]{albri}{Albright, N.: Markov formulation of the repair-misrepair model of cell survival. Radiat. Res. 118, 1-20 (1989)}

\bibitem[Almog(2010)]{almog}{Almog, N.: Molecular mechanisms underlying tumor dormancy. Cancer Lett. 294, 139-146 (2010)}

\bibitem[Anderson (2007)]{book1}{Anderson A. R., Chaplain M. A., Rejniak K. A.: Single-cell-based models in biology and medicine. Basel (Switzerland), Springer, 2007}

\bibitem[Athale et al.(2005)]{athale}{Athale, C., Mansury, Y., Deisboeck, T.S.: Simulating the impact of a molecular 'decision-process' on cellular phenotype and multicellular patterns in brain tumors. J. Theor. Biol. 233, 469-481 (2005)}

\bibitem[Azzam et al.(1998)]{azzam}{Azzam, E.I., De Toledo, S. M., Gooding, T., Little, J. B.: Intercellular communication is involved in the bystander regulation of gene expression in human cells exposed to very low fluences of alpha particles. Radiat. Res. 150, 497-504 (1998)}


\bibitem[Blyth $\&$ Sykes(2001)]{blyth}{Blyth, B.J., Sykes, P. J.: Radiation-induced bystander effects: what are they, and how relevant are they to human radiation exposures? Radiat. Res. 176, 139-157 (2001)}

\bibitem[Brenner(2008)]{brenner}{Brenner, D. J.: The linear-quadratic model is an appropriate methodology for determining iso-effective doses at large doses per fraction. Semin. Radiat. Oncol. 18, 234-239 (2008)}

\bibitem[Brunton $\&$ Wheldon(1980)]{brunton}{Brunton, G. F., Wheldon, T. E.: The Gompertz equation and the construction of tumor growth curves. Cell Tissue Kinet. 13, 455-460 (1980)}

\bibitem[Byrne et al.(2009)]{byrne}{Byrne H., Drasdo D.: Individual-based and continuum models of growing cell populations: a comparison. J. Math. Biol. 58, 657-87 (2009)}

\bibitem[Calini et al.(2015)]{calini}{Calini, V., Urani, C., Camatini, M. Comet assay evaluation of DNA single- and double-strand breaks induction and repair in C3H10T1/2 cells. Cell Biol. Toxicol. 18, 369-379 (2002)}

\bibitem[Cilfone et al.(2015)]{cilfone}{Cilfone, N. A., Kirschner, D. E., Linderman, J. J.: Strategies for efficient numerical implementation of hybrid multi-scale agent-based models to describe biological systems. Cell. Mol. Bioeng. 8, 119-136 (2015)}

\bibitem[Cucinotta et al.(2008)]{cucinotta}{Cucinotta, F. A., Pluth, J. M., Anderson, J. A., Harper, J. V., O'Neill, P.: Biochemical kinetics model of DSB repair and induction of $\gamma$-H2AX foci by non-homologous end joining. Radiat. Res. 169, 214-222 (2008)}

\bibitem[Curtis(1986)]{curtis}{Curtis, S. B.: Lethal and potentially lethal lesions induced by radiation - A unified repair model. Radiat. Res. 106, 252-270 (1986); Erratum, Radiat. Res. 119, 584 (1989).}

\bibitem[Dale(1985)]{dale}{Dale, R. G.: The application of the linear-quadratic dose-effect equation to fractionated and protracted radiotherapy. Br. J. Radiol. 58, 515-528 (1985)}

\bibitem[Davidson et al.(1986)]{david}{Davidson, T., Westbury, G., Harmer, C. L.: Radiation-induced soft-tissue sarcoma. Br. J. Surgery 73, 308-309 (1986)}

\bibitem[Deisboeck et al.(2011)]{deisboeck}{Deisboeck, T. S., Wang, Z., Macklin, P., Cristini, V.: Multiscale cancer modeling. Ann. Rev. Biomed. Eng., 13, 127-55 (2011)}

\bibitem[Demerec $\&$ Latarjet(1946)]{demerec}{Demerec, M., Latarjet, R.: Mutations in bacteria induced by radiations. Cold Spring Harbor Symp. Quant. Biol., 11, 38-50 (1946)}

\bibitem[Dikomey et al.(1998)]{diko}{Dikomey, E., Dahm-Daphi, J., Brammer, I., Martensen, R., Kaina, B.: Correlation between cellular radiosensitivity and non-repaired double-strand breaks studied in nine mammalian cell lines. Int. J. Radiat. Biol. 73, 269-278 (1998)}

\bibitem[Edelman et al.(2010)]{edelman}{Edelman, L. B.,Eddy, J. A.,Price, N. D.: In-silico models of cancer. Wiley Interdiscip. Rev. Syst. Biol. Med. 2, 438-59 (2010)}

\bibitem[El-Awady et al.(2003)]{awadi}{El-Awady, R. A., Dikomey, E., Dahm-Daphi, J.: Radiosensitivity of human tumour cells is correlated with the induction but not with the repair of DNA double-strand breaks, British J. Cancer 89, 593-601 (2003)}

\bibitem[Cristofalo et al.(2004)]{cristof}{Cristofalo, V. J., Lorenzini, A, Allen, R. G., Torres, C., Tresini, M.: Replicative senescence: a critical review. Mech. Ageing Dev., 125, 827-848 (2004)}

\bibitem[Georgescu et al.(2013)]{plosdsb}{Georgescu, W., Osseiran, A., Rojec, M., Liu, Y., Bombrun, M., Tang, J., Costes, S. V. Characterizing the DNA damage response by cell tracking algorithms and cell features classification using high-content time-lapse analysis. PLoS ONE 10, e0129438 (2015)}

\bibitem[Gerashchenko $\&$ Howell(2003)]{gerash1}{Gerashchenko, B. I., Howell, R. W.: Cell proximity is a prerequisite for the proliferative response of bystander cells co-cultured with cells irradiated with gamma-rays. Cytometry, 56A, 71-80 (2003)}

\bibitem[Gerashchenko $\&$ Howell(2005)]{gerash2}{Gerashchenko, B. I., Howell, R. W.: Bystander cell proliferation is modulated by the number of adjacent cells that were exposed to ionizing radiation. Citometry, 66A, 62-70 (2005)}

\bibitem[Gibbs(2000)]{gibbs}{Gibbs, J. B.: Mechanism-based target identification and drug discovery in cancer research. Science 287, 1969-1973 (2000)}

\bibitem[Graeber et al.(2000)]{graeber}{Graeber, T. G., Osmanian, C., Jacks, T., et al.: Hypoxia-mediated selection of cells with diminished apoptotic potential in solid tumours. Nature 379, 88-91 (1996)}

\bibitem[Hayflick(1965)]{hayflick}{Hayflick, G.: The limited \emph{in vitro} lifetime of human diploid cells strains. Exp. Cell Res. 37, 614-36 (1965)}

 \bibitem[Iyer $\&$ Lehnert(2002)]{iyer1}{Iyer, R., Lehnert, B. E.: Alpha-particle-induced increases in the radioresistance of normal human bystander cells. Radiat. Res. 157, 3-7 (2002)}
 
 \bibitem[Iyer $\&$ Lehnert(2000)]{iyer2}{Iyer, R., Lehnert, B. E.: Factors underlying the cell growth-related bystander responses to $\alpha$-particles. Cancer Res. 60, 1290-1298, (2000)}

\bibitem[Kansal et al.(2000)]{torquato}{Kansal, A. R., Torquato, S., Harsh, G. R., Chiocca, E. A., Deisboeck, T. S.: Simulated brain tumor growth dynamics using a three-dimensional cellular automaton. J. Theor. Biol. 203, 367-382 (2000)}

\bibitem[Keinj et al.(2011)]{keinj}{Keinj, R., Bastogne, T., Vallois, P.: Multinomial model-based formulations of TCP and NTCP for radiotherapy treatment planning. J. Theor. Biol. 279, 59-64 (2011)}

\bibitem[Kellerer(1996)]{keller}{Kellerer, A. M.: Fundamentals of microdosimetry. In: Kase, K., Bjarngard, B., Attix, F. (eds.) The dosimetry of ionizing radiation, pp. 78-162. Orlando (USA), Academic Press, 1985}

\bibitem[Kempf et al.(2013)]{kempf}{Kempf, H., Hatzikirou, H., Bleicher, M., Meyer-Hermann, M. In-silico analysis of cell cycle synchronisation effects in radiotherapy of tumour spheroids. PLoS Comput. Biol. 9, e1003295 (2013)}

\bibitem[Kennedy et al.(1984)]{kennedy}{Kennedy, A.R., Cairns, J., Little, J.B.: Timing of the steps in transformation of C3H 10T$\sfrac{1}{2}$ cells by X-irradiation. Nature, 307, 85-86, (1984)}

\bibitem[Kutalik et al.(1996)]{kutalik}{Kutalik, Z., Razaz, M., Baranyi, J. Connection between stochastic and deterministic modelling of microbial growth. J. Theor. Biol. 232, 285-299 (2005)} 

\bibitem[Kutcher(1996)]{kutcher}{Kutcher, G. J.: Quantitative plan evaluation: TCP/NTCP models. Front. Radiat. Ther. Oncol., 29, 67-80 (1996)}

\bibitem[Little(2003)]{little}{Little, J. B.: Genomic instability and bystander effects: a historical perspective. Oncogene, 22, 6978-6987 (2003)}

\bibitem[Lobrich et al.(1993)]{lobrich}{Lobrich, M., Ikpeme, S., Haub, P., Weber, K. J., Kiefer, J.: DNA double-strand break induction in yeast by X-rays and alpha-particles measured by pulsed-field gel electrophoresis. Int. J. Radiat. Biol. 64, 539-546 (1993)}

\bibitem[Lowengrub et al.(2010)]{lowengrub}{Lowengrub, J. S., Frieboes, H. B., Jin, F., Chuang, Y. L., Li, X., Macklin, P.: Nonlinear modelling of cancer: bridging the gap between cells and tumours. Nonlinearity, 23, R1-91 (2010)}

\bibitem[Ma et al.(2013)]{mama}{Ma, N.-Y., Tinganelli, W., Maier, A., Durante, M., Kraft-Weyrather, W.: Influence of chronic hypoxia and radiation quality on cell survival. J. Radiat. Res. 54 (Suppl.) i13-i22, 2013.}

\bibitem[Martins et al.(2007)]{martins}{Martins, M. L., Ferreira, S. C., Villeja, M. J.: Multiscale models for the growth of avascular tumors. Phys. Life Rev. 4, 128-156 (2007)}

\bibitem[Moreira $\&$ Deutsch(2002)]{moreira}{Moreira, J., Deutsch, A.: Cellular automaton models of tumor development: a critical review. Adv. Complex Syst. 5, 247-267 (2002)}

\bibitem[Mothersill $\&$ Seymour(1998)]{mother}{Mothersill, C., Seymour, C. B.: Cell-cell contact during gamma irradiation is not required to induce a bystander effect in normal human keratinocytes: Evidence for release during irradiation of a signal controlling survival into the medium. Radiat. Res. 149, 256-262 (1998)}

\bibitem[Narayan et al.(1997)]{narayan}{Narayanan, P. K., Goodwin, E. H., Lehnert, B. E.: $\alpha$-particles initiate biological production of superoxide anions and hydrogen peroxide in human cells. Cancer Res., 57, 3963-3971 (1997)} 
 
\bibitem[Narayan et al.(1999)]{naraya2}{Narayanan, P. K., LaRue, K. E. A., Goodwin, E. H., Lehnert, B. E.: Alpha particles induce the production of interleukin-8 by human cells. Radiat. Res.152, 57-63 (1999)}

\bibitem[Powathil et al.(2013)]{powa}{Powathil, G. G., Adamson, D. J., Chaplain, M. A. Towards predicting the response of
a solid tumour to chemotherapy and radiotherapy treatments: clinical insights from a computational model. PLoS Comput. Biol. 9, e1003120 (2013)}

\bibitem[Robinson et al.(1988)]{robin}{Robinson, E., Neugut, A. I., Wylie P.: Clinical aspects of post-irradiation sarcoma. J. Natl. Cancer Inst. 80, 233-240 (1988)}

\bibitem[Roos $\&$ Kaina(2006)]{roos}{Roos, W. P., Kaina, B.: DNA damage-induced cell death by apoptosis. Trends Mol. Med. 12, 440-450 (2006)}

\bibitem[Rubinow(1996)]{rubinow}{Rubinow, S. I. A maturity-time representation for cell populations. Biophys. J. 8, 1055-1073 (1968)}

\bibitem[Sachs et al.(1989)]{hanf}{Sachs, R. K., Hlatky, L., Hahnfeldt, P., Chen, P.-L.: Incorporating dose-rate effects in Markov radiation survival cell models. Radiat. Res. 124, 216-226 (1990)}

\bibitem[Sachs et al.(2001)]{sachs}{Sachs, R. K., Hlatky, L. R., Hahnfeldt, P.: Simple ODE models of tumor growth and anti-angiogenic or radiation treatment. Math. Comput. Model. 33, 1297-1305 (2001)}

\bibitem[S\'anchez-Reyes(1992)]{sanchez}{S\'anchez-Reyes, A.: A simple model of radiation action in cells based on a repair saturation mechanism. Radiat. Res. 130, 139-147 (1992)}

\bibitem[Schultz et al.(2000)]{schultz}{Schultz, L. B., Chehab, N. H., Malikzay, A., Halazonetis, T. D.: P53 binding protein 1 (53bp1) is an early participant in the cellular response to DNA double-strand breaks. J. Cell Biol. 151, 1381-1390 (2000)}

\bibitem[S$\o$rensen et al.(2011)]{brita}{S$\o$rensen, B. S., Vestergaard, A., Overgaard, J., Praestegaard, L. J.: Dependence of cell survival on instantaneous dose rate of a linear accelerator. Radioth. Oncol. 101, 223-225 (2011)}

\bibitem[Stewart(2001)]{stewart}{Stewart, R. D.: Two-lesion kinetic model of double-strand break rejoining and cell killing. Radiat. Res. 156, 365-378 (2001)}

\bibitem[Stukalin et al.(1996)]{stukalin}{Stukalin, E. B., Aifuwa, I., Kim, J. S., Wirtz, D., Sun, S. X. Age-dependent stochastic models for understanding population fluctuations in continuously cultured cells. J. Roy. Soc. Interface 10, 0325 (2013)}

\bibitem[Tracqui(2009)]{tracqui}{Tracqui, P.: Biophysical models of tumour growth. Rep. Prog. Phys. 72, 056701 (2009)}

\bibitem[Terzaghi $\&$ Little(1976)]{terzaghi}{Terzaghi M., Little J.B.: X-radiation-induced Transformation in a C3H mouse embryoderived cell line. Cancer Res., 36, 1367-1374 (1976)}

\bibitem[Tobias(1985)]{tobias}{Tobias, C. A.: The repair-misrepair model in radiobiology: comparison to other models. Radiat. Res. 104 (Suppl.), 77-95 (1985)}

\bibitem[Wang et al.(2007)]{wang2}{Wang, Z., Zhang, L., Sagotsky, J., Deisboeck, T. S.: Simulating non-small cell lung cancer with a multiscale agent-based model. Theor. Biol. Med. Mod. 4, 50 (2007)}

\bibitem[Wang et al.(2015)]{wang}{Wang, Z., Butner, J. D., Kerketta, R., Cristini, V., Deisboeck, T. S.: Simulating cancer growth with multiscale agent-based modeling. Semin. Cancer Biol. 30, 70-78 (2015)}

\bibitem[Wei et al.(2013)]{wei}{Wei, L., Nakajima, S., Hsieh, C. L., Kanno, S., Masutani, M., Levine, A. S., Yasui, A., Lan, L.: Damage response of XRCC1 at sites of DNA single strand breaks. J. Cell Sci. 126, 4414-4423 (2013)}

\bibitem[Wlodek $\&$ Hittelman(1988)]{wlodek}{Wlodek D, Hittelman WN. The relationship of DNA and chromosome damage to survival of synchronized X-irradiated L5178Y cells: II. Repair. Radiat. Res. 115, 566-575 (1988)}

\bibitem[Yachida et al.(2010)]{yachi}{Yachida, S., Jones, S., Bozic, I., Antal, T., Leary, R.: Distant metastasis occurs late during the genetic evolution of pancreatic cancer. Nature 467, 1114-1117 (2010)}





\end{thebibliography}
\end{document}